\documentstyle[aps,tighten,12pt,epsfig]{revtex}
\newcommand {\bi} {\bibitem}
\newcommand {\be} {\begin{equation}}
\newcommand {\beq} {\begin{eqnarray} \nonumber }
\newcommand {\ee} {\end{equation}}
\newcommand {\eps} {\epsilon}
\newcommand {\ei} {\epsilon_i}
\newcommand {\ej} {\epsilon_j}
\newcommand {\ek} {\epsilon_k}
\newcommand {\epp} {\epsilon^{\prime}}
\newcommand {\si} {\sigma}

\newcommand {\al} {\alpha}
\newcommand {\aeps} {\alpha_{\epsilon}}

\newcommand {\app} {\alpha_+}
\newcommand {\amm} {\alpha_-}

\newcommand {\Tr} {\mbox{Tr}}

\newcommand{\bq}{\begin{eqnarray}}
\newcommand{\eq}{\end{eqnarray}}
\newcommand{\n}{\noindent}
\newcommand{\nn}{\nonumber\\}
\newcommand{\siml}{\stackrel{<}{\sim}}
\newcommand{\simg}{\stackrel{>}{\sim}}
\newcommand{\bc}{\begin{center}}
\newcommand{\ec}{\end{center}}                                   

\newcommand {\s} {\sigma}

\newcommand {\w} {\omega}

\def\(({\left(}
\def\)){\right)}
\def\[[{\left[}
\def\]]{\right]}
\def\bi{\bibitem}

\def \(({\left(}
\def \)){\right)}
\def \[[{\left[}
\def \]]{\right]}

\def \nn{\nonumber}

\def \tk{T_K}
\def \teff{T_{eff}}
\newcommand {\la} {\langle}
\newcommand {\ra} {\rangle}

\begin{document}

\title{Thermodynamics of binary mixture glasses}
\author{Barbara Coluzzi$^{a}$, 
Marc M\'ezard$^{b}$, Giorgio Parisi$^{c}$ and Paolo Verrocchio$^{c}$}

\maketitle

\begin{center}
{
a) John von Neumann-Institut f\"ur Computing (NIC) \\
c/o Forschungszentrum J\"ulich \\
D-52425 J\"ulich (Germany) \\
b) Institute for Theoretical Physics \\
{University of California Santa Barbara, CA 93106-4030, (USA)} \\ 
{and Physique Th\'eorique-ENS, CNRS, Paris (France)} \\
c) Dipartimento di Fisica and Sezione INFN,\\
Universit\`a di Roma ``La Sapienza'',
Piazzale Aldo Moro 2,
I-00185 Rome (Italy)
}
\end{center}

\date{\today}

\vspace{2cm}

\begin{abstract}

\noindent

We compute the thermodynamic properties of the glass phase in a binary 
mixture of soft spheres. Our approach is a generalization to mixtures of the
replica strategy, recently proposed by Mezard and Parisi,
providing
a first principle statistical mechanics computation of the thermodynamics
of glasses.
The method starts from the inter-atomic potentials, and
translates the problem into the study of a molecular liquid.
We compare our analytical predictions to numerical simulations,
focusing  onto the values of the thermodynamic transition temperature
and the configurational entropy. 
\end{abstract}


\newpage

\begin{section}{Introduction}
\noindent
In this paper we present the generalization to the binary
mixture case of a thermodynamic theory of glasses, recently proposed in
\cite{MePa1,MePa2}, which allows to deduce equilibrium properties of fragile 
glasses
\cite{An} from those of the corresponding liquid phase, 
computed for a molecular
liquid consisting of $m$ 'clones' \cite{Me} of the system with $m < 1$.

The hypothesis at the heart of this strategy is the existence of a 
liquid-glass thermodynamic transition, driven by the 
'entropy crisis' predicted by Kauzmann \cite{Ka}, 
and the scenario is similar to the 
one described by Adam, Gibbs and Di Marzio \cite{AdGi,GiDi,Pa1}. 
The transition considered here can be also explained in terms of
a certain type of replica symmetry breaking
 (called `one step replica symmetry 
breaking' -1RSB). It shares its main features with 
the glass transition found in some 
discontinuous spin-glasses model,
as first proposed by Kirkpatrick Thirumalai and Wolynes \cite{KiThWo}.

We identify the mode coupling temperature $T_{MCT}$ \cite{GoSj}
with the dynamical temperature $T_{D}$ 
which exists in discontinuous  spin-glasses \cite{Horner,CuKu,BoCuKuMe}, 
and we assume that below this temperature
the phase space can be partitioned in a very large number of different
free energy valleys. 
These valleys are supposed to be, in terms of free-energy, the
equivalent of the so called inherent structures \cite{St}, 
which are built from
the minima of the potential energy together with their basins of attraction.

In other words, we suppose that, for $T<T_{MCT}$, 
a typical equilibrium configuration belongs to one of these valleys.
We label the valleys with an index $\al$, 
and denote for each valley the  free energy density as $f_{\al}$,
the subset of equilibrium configurations belonging to
the valley as ${\cal V}_{\alpha}$ and the corresponding restricted partition
function as $Z_{\al}$.
The canonical partition function can then be written in the following way:

\bq
\label{partizione}
Z & \simeq & \sum_{\al} Z_{\al} \ = \ 
\sum_{\al} \ \int_{\{ x \} \in {\cal V}_{\alpha}} d x \  e^{- \beta H(x)} \ = \ 
\sum_{\al} e^{- N \beta f_{\al}} 
\eq
where the function $H$ is the Hamiltonian of the system and $\beta$ 
is the inverse temperature. 
The number of valleys with a given value of free energy density is defined as 
${\cal N}(f) \equiv \sum_{\al} \delta (f - f_{\al})$, and we 
assume that in the thermodynamic limit it becomes a continuous
function. It is then possible to write the partition function as:

\bq
Z  & \simeq & \int df \  {\cal N}(f) e^{- N \beta f} \ = \
\int df \ e^{- N \left[ \beta f - \Sigma(f,T) \right] }
\eq
where we have  introduced the
complexity $\Sigma \equiv \log {\cal N}/N$.

Let us note that the system in equilibrium does not minimize 
the free energy of the single valleys,
but a 'collective' thermodynamic potential $\phi(T)$,
that we interpret as the actual free energy in the liquid and glass phases.
$\phi(T)$ is defined by:

\be
\label{potgener}
\phi(T) \equiv f^* - T \Sigma(f^*,T).
\ee
where $f^*$ is the temperature dependent free energy which minimizes the function
$f - T \Sigma(f,T)$.

In this picture,
the total entropy density is the sum of   
the entropy inside the valley, and of the entropy coming from the
very large number of valleys, with the same value of free energy,
that the system is allowed to explore:
\be
S_{tot} = S_{valley} + \Sigma(T,f^*) \ ,
\ee
hence the complexity introduced here is completely equivalent to the
usual concept of configurational entropy of a super-cooled liquid.

Assuming the existence of this decomposition of phase space
into valleys,
we will show that there is a finite temperature $\tk$ (the so-called
Kauzmann temperature) where the system undergoes a 
thermodynamic transition with the following features:

\begin{itemize} 

\item 
$\tk$ is the temperature where the complexity $\Sigma$ vanishes 
\cite{MePa1,MePa2}. This means that, unlike the liquid phase, 
in the whole low temperature glass phase, 
only a non exponential number of valleys contribute to the partition function, 
namely the ones with the lowest free energy density $f_{min}$.

\item 
At $\tk$ there is a second order transition from the thermodynamic point 
of view. The free energy is continuous and there is no latent heat. 
The specific heat 'jumps' from the liquid value to a smaller one,
in agreement with the Dulong and Petit law. 

\item
At $\tk$ there is a discontinuity of the order parameter. 
Below $\tk$, in the glass phase, the system is an amorphous solid 
and the thermal average of the local particle density becomes non uniform,
exhibiting peaks at the favoured positions where the particles 
tend to be trapped in some cages. The order parameter is related
to the spatial modulation of the density, and it goes 
discontinuously from zero in the high temperature liquid phase to
a finite modulation in the glass phase.

\end{itemize}

This transition  could
be experimentally observed only if one would be able to cool the liquid 
at an infinitely slow rate, and $\tk$ should correspond to the
temperature where the viscosity is supposed to diverge 
(following for instance a
generalized Vogel-Fulcher law $\eta \propto \exp (T-T_k)^{-\nu}$) \cite{An}.
In real experiments, infinitely slow cooling is not available, and
the correlation time becomes of order of the experimental time at a
temperature $T_g$, which is  in general 
an intermediate temperature $\tk < T_{g} < T_{MCT}$.
The value of $T_g$ could be computed only if we had under control the
time dependence of the correlation functions. In this paper  we
study only static quantities, and we cannot say anything about
the value of $T_g$ or the temperature dependence of the viscosity above $T_K$.

In such an 'entropy crisis' scenario, it has been 
shown \cite{Mo,Me} that the thermodynamic properties of the glass phase 
can be computed in principle by considering $m$ replicas of
the  original system, constrained to stay in the same valley, 
by means of a small but extensive coupling term. In this case, the
arguments used in the derivation of (\ref{potgener}) can be applied again, leading to
a replicated version of the same equation:

\be
\label{replicato}
\Phi(m,T) \: \equiv Min|_f \: \(( m \: f - T \Sigma(f,T)\)) \ .
\ee
Once again, each of the $m$ systems does not reach the lowest possible 
free energy, but the one which optimizes the balance between the free energy and
complexity in (\ref{replicato}).

Interestingly enough, 
one can derive many properties of the system from (\ref{replicato})
if one is able to continue it analytically  and compute it 
for any real value of $m$,  thinking about $m$ as
a new parameter of the problem. Indeed,  $\Phi(m,T)$,
considered as a function of $m$, gives access to the configurational
entropy (complexity) $\Sigma(f,T)$ through a Legendre transform. 
This implies the relations:

\be
\label{emmesigma}
f = {\partial \Phi(m,T) \over \partial m}  \hspace{.5cm}
\Sigma  = {m^2 \over T} {\partial (\Phi(m,T)/m) \over \partial m}.
\ee
from which it is possible to eliminate $m$, obtaining
$\Sigma(f)$, which measure the number of valleys with a given value of free 
energy $f$.
Let us underline that (\ref{emmesigma}) gives access to the full
curve of complexity versus  free energy, while the equilibrium 
free energy
of the physical system is obtained only after taking the limit $m \to 1$.

As we shall see, the thermodynamic potential $\Phi(m)/m$ is a convex 
function of $m$ with a maximum at a point $m^*(T)$, which is
an increasing function of $T$, vanishing at $T=0$.
The second equation of (\ref{emmesigma}) is thus well-defined
for  $m \le m^*(T)$. At $m=m^*(T)$, the resulting complexity
vanishes  $\Sigma =0$ and 
the free energy $\partial \Phi /\partial m$ reaches a value $ f_{min}$. 
For $m<m^*(T)$ the complexity $\Sigma$ is non zero: it is thermodynamically
favourable to select some valleys which have a free energy density
larger than $f_{min}$, because of the corresponding gain
in complexity.
If one increases $m$ beyond $m^*(T)$, the formula (\ref{emmesigma})
gives an unphysical negative complexity. In fact in the whole region
$m>m^*(T)$ the correct value of $f$ is $f=f_{min}$, and the
complexity is zero.
 
This is easily understood from the physical interpretation
of the transition
which we  now turn to.
The above scenario has:
\begin{itemize}
\item 
a high temperature phase where $m^*(T)>1$.  In this phase, when the 
limit $m \to 1$ is performed and the equilibrium free energy of the 
original system is recovered, one gets a value $f_{eq} > f_{min}$
together
with a positive configurational entropy.
\item
a low temperature phase where $m^*(T) <1$. In this phase,
in the limit $m \to 1$ the
equilibrium free energy is $f_{min}$ and the configurational entropy
is null.
\end{itemize}
It is quite easy, at this point, to recognize these two 
thermodynamic phases as the super-cooled liquid one (high) and the 
glass one (low), separated by a thermodynamic transition of 
second order, driven by the vanishing of the configurational
entropy, at the temperature $\tk$ where $m^*(\tk) =1$. 
All the thermodynamic quantities in the glass phase
can be computed from the replicated free energy (\ref{replicato}) at
the point $m^*$, which play the role of the free energy of the glass.
 
This scenario of the glass transition
is identical to the phase transition appearing
in discontinuous (1RSB) spin glasses where it was first explained
\cite{MePaVi}. The simplest example of such a
discontinuous spin glass transition is the Random Energy Model
\cite{REM} which displays a total freezing at $T=T_K$.
As first noticed by Kirkpatrick Thirumalai and Wolynes
\cite{KiThWo}, discontinuous spin glasses provide 
some well defined mean field systems where
 the old ideas of Adam-Gibbs-Di Marzio of a real 
thermodynamic transition driven by entropic reasons are at work.
The present approach allows to apply the replica method 
directly to the structural glasses (in spite of the absence of
any quenched disorder in the Hamiltonian). Assuming that the
structural glass transition is characterized in the replica language 
by a 1RSB, as in discontinuous spin glasses \cite{CoPa,BaKo,Pa2},
we can compute the thermodynamic properties of the glass phase.
The comparison  with the numerical results allows then to justify a posteriori
the main hypothesis.

At this stage, $m$ appears as an auxiliary parameter which may be interpreted as the effecive 
temperature of the vallyes; moreover one finds that in 
the low temperature phase $1-m$ 
gives the probability of finding two
systems in the same valley\cite{MePaVi}.

Summarizing, the study of the liquid-glass transition and the investigation
of the low temperature phase can be accomplished by 
computing the free energy of a replicated system
in its liquid phase, or in other
words \cite{Me}, 
the free energy of a molecular liquid where each molecule has $m$ 
atoms.
The thermodynamic properties of the glass phase can be deduced by means of the
analytic continuation to arbitrary real values of this parameter.

In the previous works \cite{MePa1,MePa2} this general approach was 
applied to a pure soft sphere system.
The extension to binary mixtures is particularly important since there are well 
known examples 
of glass forming binary mixtures where an appropriate choice of the interaction 
parameters
strongly inhibits crystalization. This allows therefore to get numerical results 
which can be compared to
the analytical ones. Here we will consider in particular a mixture of
soft spheres.

After discussing the model in sect.II, we will present in sect.III
the generalization to binary mixtures both of the 
small cage expansion and of the harmonic re-summation scheme introduced
previously \cite{MePa1} to deal with the molecular fluid. Sect.IV describes
the application of the HNC approximation to the center of mass degrees of freedom
of the molecular fluid. In the 
last section
we will discuss our analytic results, together with  some
strategies for  evaluating numerically the glass transition 
temperature $T_K$ and the
configurational entropy behavior, and a comparison between 
the numerical estimates 
and those obtained analytically.

\end{section}

\begin{section}{General framework}
We study mixtures composed of two types of particles 
called $+$ and $-$, with pairwise interactions.
The Hamiltonian of our problem is:

\be
H=\sum_{1 \le i \leq j \le N} V^{\epsilon_i \epsilon_j}(x_i-x_j) \ \ \ \ 
\epsilon_i \in \{-,+\},
\label{zn1}
\ee 

\noindent
where the $N$ particles move in a volume $V$ of a $d$-dimensional space, 
and $V^{++},V^{+-}$,  $V^{--}$ are arbitrary short range interaction 
potentials. We call $c_+$ (resp. $c_-$) the fraction of $+$ (resp. $-$) 
particles.

In the explicit computations described in the next section,  
we have  chosen a  binary mixture of soft spheres that has been extensively
 studied in the past through numerical simulations
\cite{CoPa,Han2}. The potentials are given by:
\be
V^{\eps\epp}(r)= \left ( \frac{\sigma_{\eps \epp}}{r} \right )^{12},
\ee
where
\be
\hspace{.3in} \frac{\s_{++}}{\s_{--}}=1.2, \hspace{.2in} 
\s_{+-}=\frac{\s_{++}+\s_{--}}{2} \ .
\label{pot}
\ee
The concentration is taken as $c_+=1/2$, and
the choice of the ratio $R\equiv\s_{++}/\s_{--}=1.2$ is known to strongly inhibit
crystalization.  
We also make the usual choice of considering particles with
average diameter 1 by setting
\begin{equation}
\frac{(\sigma_{++})^3+2(\sigma_{+-})^3+(\sigma_{--})^3}{4}=1.
\end{equation} 

\noindent
All thermodynamic quantities depend on the
density $\rho=N/V$ and temperature $T$ only 
through the parameter $\Gamma \equiv
\rho T^{-1/4}$. For $\Gamma$ larger than $\Gamma_D = 1.45$ (corresponding to
lower temperatures) the dynamics becomes
very slow and the autocorrelation time is very large.  Hence the system
enters the 'aging' regime, where violations of the
equilibrium fluctuation-dissipation theorem are observed \cite{Pa2}.  
This value of $\Gamma_D$ is supposed to correspond to the
mode coupling transition 
below which the relaxation is dominated by 
activated
processes \cite{Han2}.  If this simple model behaves like a real fragile
glass the Kauzmann transition, characterized 
by a discontinuity in the specific heat, is located below 
the dynamical transition, and  cannot be directly accessed 
by numerical simulations, maybe with the exception of studies done on
very small samples \cite{CoPa}.

The application of the theory to a more realistic potential, namely a Lennard-Jones 
binary mixture, 
will be treated in detail in a forthcoming paper \cite{CoPaVe}. 

As previously explained, in order to obtain some information about the super-cooled 
liquid-glass 
thermodynamic transition, we consider the thermodynamics of a molecular liquid, 
whose molecules 
are composed of $m$ atoms, each carrying a different replica index.  The tendency 
to form molecules is 
forced by a small but extensive coupling term between particles of different 
replicas \cite{Me}.  
Unlike the pure case, we are dealing here
with a situation where particles are not 
all 
indistinguishable:
we have particles of the "$+$" type and of the "$-$" type.
Physically this has an important effect when $R$ is not close to one.
At $R \simeq 1$, it is clear that the valleys of the mixture
are close to those of the pure system. More precisely, taking one 
given valley of the pure ($R=1$) system, one can generate
$N!/N_+! \: N_-!$
valleys of the mixture with $R \simeq 1$, by choosing at random
the positions of the $+$ and the $-$ particles: in this limit the
main effect of the mixture is to add a factor to the entropy, whose 
value is  $N \log 2$ when $N_{\pm} = N/2$.
On the other hand, when $R$ is very different from one, 
the valleys of the mixture are very different from those of the 
pure system; one cannot find a new valley by just exchanging a 
$+$ particle with a $-$ particle. This physical problem has an exact 
counterpart in replica space.
  One could study the case where  molecules are formed by one 
  particle of each of the $m$ 
different replicas, 
irrespective of their $\pm$ nature.  Qualitatively speaking, this would mean that 
interchanging two 
particles of different types, the two replicas 
to which particles belong would remain in 
the same valley, that is their free-energy would not change.  
There are two extreme possibilities, corresponding to the
two cases discussed above:  

\begin{itemize}
\item For $R$ very near to one, the system behaves  similarly to the  system at 
$R=1$. One can form molecules with particles of any type, and the
exchange of a "$+$" particle with a "$-$" one
 gives a very small change in free energy.
\item For $R$ quite different from 1, the exchange  of a "$+$" particle 
with a "$-$" one is a process that can be safely 
neglected, since it gives a 
variation in energy that is much larger than $kT$.  In this second case
the molecules are built up of atoms of the same type.
\end{itemize}

In each of these extreme  cases the computation is simple: 
in the first case it just reduces to 
the $R=1$ computation.
In the second case, we can assume, as we shall do here, that each molecule 
is built from $m$ atoms which are all of the same type (all "$+$" or all 
"$-$").
Then one only needs considering attractive 
coupling terms only between particles of the same kind.  The computations in the 
crossover region 
are rather complex.  For our case $R=1.2$, we have decided to neglect
 this kind of corrections and to 
consider the molecules consisting only of particles of the same type.

The replicated partition function is:

\begin{eqnarray}
\label{zm}
Z_m[\w] & = & 
\frac{1}{N_{+}!^m N_{-}!^m} 
\sum_{\si_a} \sum_{\pi_a} \int \prod_{a} 
d^d x_i^a\ 
\exp 
\left ( - \frac{\beta}{2} \sum_{i \neq j,a} V ^{\eps_i \eps_j} (x_i^a - x_j^a)
 + \right. \nonumber  \\ 
& - & \left. \sum_{i \in \{+\}} \sum_{a \neq b} 
{\w}_+(x_{\si_a (i)}-x_{\si_b (i)}) - 
\sum_{i \in \{-\}} \sum_{a \neq b} 
{\w}_-(x_{\pi_a (i)}-x_{\pi_b (i)})
\right ),
\end{eqnarray}

\noindent
where the sum over permutations of atoms in each molecule is taken into account,
and $N_{+}=c_+ N$, $N_{-}=c_- N$.
When relabeling particles, so that  particle
$i$ of a given type in replica $a$ corresponds to particle $i$ of the same
type in replica $b$ (which is supposed to belong to the same molecules) and so 
on, the sum over permutations gives a factor $(N_+! \: N_-!)^{(m-1)}$.

As discussed in the previous section, in the glass
phase the replicas becomes correlated, so 
the study of the transition is accomplished by choosing as order parameters 
the $m$-points correlation functions for each of
the two different types of particles:

\be
\label{ord1}
\rho_+(r^1,...,r^m)= \sum_{i \in \{+\}} <\delta(x_i^1-r^1)...\delta(x_i^m-r^m)>\ ,
\ee 

\be
\label{ord2}
\rho_-(r^1,...,r^m)= \sum_{i \in \{-\}} <\delta(x_i^1-r^1)...\delta(x_i^m-r^m)> \ .
\ee 

\noindent
 The transition is signaled, then, by 
the onset of an off-diagonal non trivial correlation in replica 
space at $T_K$, when the coupling functions $\omega_{\pm}$ are sent to zero. 
This feature is studied as usual  introducing the Legendre 
transform of the molecular (replicated) free energy:

\bq
\label{gdia}
G[\rho] & = &  \mathop{\rm lim}_
{ 
\begin{array}{c}
{\small m  \to 1}  \\  
{\small \w \to 0} \\
{\small N  \to \infty}
\end{array}
}
-{1 \over \beta m} \log Z_{m}[\w] - {1 \over m} 
\int \prod_{a=1}^m d^d r^a \sum_{\eps = +,-} \rho_{\eps}(r^1,...,r^m) 
W_{\eps}(r^1,...,r^m)
\eq
with
\be
W_{\eps}(r^1,...,r^m) = \sum_{a < b} \w_{\eps}(r^a-r^b)
\ee

\noindent
Performing the limit $\omega_\pm \to 0$ is equivalent
to searching  a saddle point of the functional $G[\rho]$.
In the presence of a glassy transition we expect the
following behavior
of order parameters and thermodynamic quantities:

\begin{itemize}
\item For $T > \tk$ the free energy is the liquid one ($m=1$) and the order 
parameters are trivial, i.e. $\rho_\pm(r^1)
=c_{\pm} \rho$.

\item For $T < \tk$, the glass free energy is the maximum 
with respect to $m$ of the
replicated free energy, and this maximum  is found at $m^* <1$.
The correlations $\rho_{\pm}$ become non trivial.
 From the free energy at the maximum
  we can compute all the thermodynamic quantities. 
\end{itemize}
The free energy and his first derivatives are continuous at $T_K$, while the 
heat capacity falls suddenly from liquid-like to solid-like values
when the temperature is decreased through $\tk$.
The transition, then, is of second order from the point of view of
thermodynamics, but it is discontinuous in the order parameter which
abruptly becomes a non trivial function of positions in different replicas.

It is natural to describe the particle positions in term of center of
mass coordinates $r_i$ and relative displacements $u_i^a$ with 
$x_i^a = z_i + u_i^a$ and $\sum_a u_i^a =0$.
A useful simplification is the choice, for the
polarising potentials  $\omega_\pm$,
of a quadratic coupling 
that allows to rewrite (\ref{zm}) as:

\bq
Z_m & = & \frac{1}{N^+! \: N^-!} \int \left ( \prod_{i=1}^N d^d z_i \right )
\left (\prod_{a=1}^m \: \prod_{i=1}^N d^d u^a_i \right )
 \left [ \prod_{i=1}^N \left ( m^d \delta ( \sum_{a=1}^m u^a_i) \right ) 
\right ] \cdot \nonumber \\
 & \cdot & \exp  \left ( -\beta \sum_{a=1}^m 
\sum_{i < j } V^{\al_i \beta_j}(z_i-z_j+ u^a_i-u^a_j) \right. 
+ \nn \\
& - & \left. \frac{1}{4 \app} \sum_{a, b } \sum_{i \in +} (u^a_i - u^b_i)^2
-\frac{1}{4 \amm} \sum_{a, b } \sum_{i \in -} (u^a_i - u^b_i)^2
\right ) \ .
\label{zmquad}
\eq
In the absence of the interacting potential $V$, the $\{ u^a_{i\mu}\}$ for a 
given $i$ are Gaussian
random variables with a vanishing first moment and a second moment given by
\be
\langle u^a_{i\mu} \: u^b_{i \nu} \rangle =
\left ( \delta^{ab}-\frac{1}{m} \right ) \delta_{\mu \nu} \delta_{ij}
\frac{\al_{\ei}}{m}.
\ee

\end{section}

\begin{section}{Replicated free-energy}

\begin{subsection}{Harmonic re-summation}

\n
We are interested in the regime of low temperatures, where the molecules 
are expected to have a small radius, justifying a quadratic expansion of $V$
in the partition function (\ref{zmquad}).
After integrating over these quadratic fluctuations, one obtains:
\bq
Z_m= {m^{Nd/2} \sqrt{2 \pi}^{N d (m-1)} \over {N^+! \: N^-!}} 
\int \prod_{i=1}^N d^d z_i \exp\((-\beta m \sum_{i<j}
V^{\al_i \beta_j}(z_i-z_j) -{m-1 \over 2} Tr \log \((\beta M  \)) \))
\label{zmharm}
\eq
where the matrix $M$, of dimension $Nd \times Nd$, is given by:
\be
\label{matri}
M^{\ei \ej}_{(i \mu) (j \nu)}= \delta_{ij} 
\left( \sum_k V^{\ei \ek}_{\mu\nu}(z_i-z_k) + \frac{m}{\al_{\ei}} \right)
-V^{\ei \ej}_{\mu\nu}(z_i-z_j)
\ee
and $v_{\mu\nu}(r) =\partial^2 v /\partial r_\mu \partial r_\nu$ 
(the indices $\mu$ and $\nu$, running from $1$ to $d$, denote space directions).
We have thus found an effective Hamiltonian for the centers of masses $z_i$ of 
the 
molecules, which basically looks like the original problem at the effective
 temperature $T^*=1/(\beta m)$, complicated by the contribution of 
vibration modes. We shall proceed by using the same set of
approximations which was proposed in the previous papers \cite{MePa1,MePa2}.
 We first perform a 'quenched approximation', which amounts to
 neglecting the feedback of
vibration modes onto the centers of masses, 
substituting thus
the term $Tr \log \((\beta M  \))$ in (\ref{zmharm}) by its expectation value,
for center of mass positions $z_i$ equilibrated at the temperature $T^*$. 
This approximation becomes
 exact close to the Kauzmann temperature where $m \to 1$.

Let us introduce the mean values of the diagonal terms of the matrix $M$:

\be
\label{norma}
r_{\eps} =  \sum_{\epp} c_{\eps} \rho \int{d^d r g^{\eps \epp}(r) \frac{1}{d} 
\Delta V^{\eps \epp} } + \frac{m}{\al_{\eps}}\ ,
\ee

\noindent
where 
the $g^{\eps \epp}(r)$ are the pair correlation functions. 
We neglect the fluctuation of these diagonal terms
(an approximation which should be valid at high densities) 
and normalize the off diagonal matrix elements as follows:

\be
\label{ele}
C^{\eps \epp}_{(i\mu)(j\nu)} \equiv 
\sqrt{c_{\eps} c_{\epp} \over r_{\eps} r_{\epp}} V^{\eps \epp}(z_i-z_j).
\ee

\noindent
The replicated free energy per particle, 
$\phi(m,T) \equiv \Phi(m,T)/m$, can be expanded in series:

\bq
\label{fharmoin}
{\beta \phi(m,\beta)} & = & 
-{d \over 2 m} \log(m)- { d (m-1) \over 2 m } \log(2 \pi)
-{1 \over m N} \log Z_{liq}(\beta \: m) + \nonumber \\  
& + & \frac{d \:(m-1)}{2m} \left (c_+ \: \log(\beta \: r_+) 
+c_- \log(\beta \: r_-) \right ) + \nn \\
& + & \frac{1}{N} {(m-1) \over 2 m}  \sum_{p=2}^{\infty}
 \left \langle \frac{\Tr C^p}{p}  \right \rangle \ ,
\eq

\noindent
where the $p$-th order term depends as usual on the $p$-points correlation
function

\bq
\left \langle Tr C^p \right \rangle & = & \sum_{\eps_1...\eps_p \in \{+,-\}} 
\sum_{\mu_1...\mu_p}
 \int d^d z_1 \dots d^d z_p \: \rho^p 
g^{\eps_1 \dots \eps_p}(z_1 \dots z_p) C^{\eps_1\eps_2}_{\mu_1\mu_2}(z_1-z_2) 
\cdots \nn \\
  & \cdots & C^{\eps_{p-1}\eps_p}_{\mu_{p-1}\mu_p}(z_{p-1}-z_p)
  C^{\eps_p \eps_1}_{\mu_p\mu_1}(z_p-z_1),
\eq

\noindent
We use a 'chain' approximation in the computations of traces,
where   terms with two equal indices are neglected ,
and the so called superposition approximation for the $p$-points correlation 
functions $g^{(p)}(z_1...z_p)=g(z_1-z_2) \cdots g(z_p-z_1)$.
With these hypotheses we arrive at:

\bq
\left \langle \Tr C^p \right \rangle & = &  
\int d^d z_1 \dots d^d z_p \: \rho^p \:  \sum_{\mu_1 \dots \mu_p} 
\sum_{\eps_1...\eps_p} 
g^{\eps_1\eps_2}(z_1-z_2)   C^{\eps_1\eps_2}_{\mu_1 \mu_2}(z_1-z_2)
\cdots \nn \\
& \cdots & g^{\eps_{p-1}\eps_p}(z_{p-1}-z_p ) C^{\eps_{p-1}\eps_p}_{\mu_{p-1} 
\mu_p}(z_{p-1}-z_p )
g^{\eps_p\eps_1}(z_p-z_1) C^{\eps_p\eps_1}_{\mu_p \mu_1}(z_p-z_1).
\eq

\noindent
The convolutions are computed in Fourier space, introducing the tensor:

\be
D_{\mu\nu}^{\eps \epp} (k) \equiv 
\int d^d r 
g^{\eps \epp} (r) C^{\eps \epp}_{\mu\nu}(r) e^{i k r}, 
\label{defab}
\ee

\noindent
which can be decomposed into its diagonal (longitudinal) and 
traceless (transversal) parts with respect to the spatial ($\mu,\nu$)
indices:

\be
D_{\mu\nu}^{\eps \epp} (k)=\delta_{\mu\nu} \ a^{\eps\epp}(k) +
\(( {k_\mu k_\nu \over k^2} -{ \delta_{\mu \nu}\over d} \)) 
b^{\eps \epp}(k).
\ee

\noindent
The last step consists in the diagonalization of $D$ in the space of the 
particles types ($\eps,\epp$). 
 For each $k$, there are four distinct eigenvalues, 
 the two `longitudinal' ones, corresponding to 
that of the $2 \times 2$ matrix

\be
D^{\eps\epp}_{\parallel}(k)=a^{\eps \epp}(k)+ {d-1 \over d} b^{\eps \epp}(k),
\ee

\noindent
and the two `transverse' eigenvalues of the matrix

\be
D^{\eps \epp}_{\perp}(k)=a^{\eps \epp}(k)-{1 \over d} b^{\eps \epp}(k).
\ee

\noindent
The eigenvalues are:

\bq
\lambda_\parallel&=&{1 \over 2} \((D^{++}_{\parallel}+D^{--}_{\parallel} + 
\sqrt{
(D^{++}_{\parallel}-D^{--}_{\parallel})^2+4 (D^{+-}_{\parallel})^2} \))
\nn \\
\mu_\parallel&=&{1 \over 2} \((D^{++}_{\parallel}+D^{--}_{\parallel} - \sqrt{
(D^{++}_{\parallel}-D^{--}_{\parallel})^2+4 (D^{+-}_{\parallel})^2}\))
\nn \\
\lambda_\perp&=&{1 \over 2} \((D^{++}_{\perp}+D^{--}_{\perp} + \sqrt{
(D^{++}_{\perp}-D^{--}_{\perp})^2+4 (D^{+-}_{\perp})^2} \))
\nn \\
\mu_\perp&=&{1 \over 2} \((D^{++}_{\perp}+D^{--}_{\perp} - \sqrt{
(D^{++}_{\perp}-D^{--}_{\perp})^2+4 (D^{+-}_{\perp})^2}\))
\eq

\n
Using these approximations, 
the expression of the binary mixture free energy per particle is:

\bq
\label{fharmofin}
\phi(m,\beta) & = & -{d \over 2 m} \log(m)- { d (m-1) \over 2 m } \log(2 \pi)
 + {d (m-1) \over 2 m } \((c_+\log (\beta r_+) + c_-\log(\beta r_-) \))
+ \nn \\
& + &
{(m-1) \over 2 m } \frac{1}{\rho} 
\int d^d k  \left\{ L_3 (\lambda_\parallel(k)) + L_3 (\mu_\parallel(k))
+  
{(d-1)} \left[ L_3 (\lambda_\perp(k)) + L_3 (\mu_\perp(k))\right] \right\} 
+ \nn \\ 
&-&
{(m-1) \over 4 m } \int d^d r \: \rho \sum_{\eps \epp}
g^{\eps \epp}(r) \sum_{\mu\nu} \((C^{\eps \epp}_{\mu\nu}(r)\))^2
-{1 \over m N} \log Z_{liq}(\beta \: m), 
\label{chain}
\eq

\noindent
where the function $L_3$ is $\log(1-x)+x+{x^2/2}$.

\n
Let us notice that the condition
for identifying the Kauzmann temperature, $ {\partial \beta F_m \over \partial m} 
|_{m=1}=0$,
reads in our harmonic approximation:

\be
S_{liq}={d \over 2} \log(2 \pi e) - {1\over 2}\la Tr \log\((\beta M  \))\ra
\ee

\n
$S_{liq}$ is the entropy of the liquid at the effective temperature
$T_{eff}$, which is equal to $T$ for $m=1$. The right hand side of 
this equation is nothing but the entropy $S_{sol}$ of an
harmonic solid with a matrix of second derivatives given by $M$. 
Thus, we find:
\be
\Sigma(\beta) = 
m^2 {\partial \beta F_m \over \partial m }{\Biggr |}_{m=1}=S_{liq}-S_{sol}
\label{comple}
\ee
If $S_{liq}<S_{sol}$, the system is  in the glassy phase ($T<T_K$), while in the other 
case $S_{liq}>S_{sol}$, 
the temperature is greater than $T_K$ (and of course less than $T_D$
if the spectrum of $M$ is positive). The complexity is then 
$\Sigma =S_{liq}-S_{sol}$, as expected on general grounds \cite{Mo}.

Formula (\ref{fharmofin}) allows to compute the free energy $\Phi(m,T)=m\phi(m,T)$
which is the main quantity needed
 to investigate the thermodynamics of the low 
temperature glass phase, using (\ref{emmesigma}).  
It should be emphasized that within the approximations
we used here, the only properties of the liquid phase
which are needed to get $\Phi$ are the pair correlation  $g(r)$ and the free-energy.
 Beside usual thermodynamic quantities (energy, entropy, heat capacity...), 
we are interested in 
the two new parameters describing the glassy phase:

\begin{itemize}

\item
The square cage radii $A_{\eps}$, defined as $A_{\eps}= {1 \over 3} (
\la x_i^2\ra -\la x_i \ra^2)$ for type $\eps$ particles. 
This square cage radii are obtained by differentiating the free energy
with respect to coupling terms and by sending couplings to zero in the end:

\be
A_{\eps} = {2 \over d (m-1) N_{\eps}} \  {\partial (\beta F) \over 
\partial (1/ \aeps)} (\aeps=\infty)
\label{gabbie}
\ee

\n
The square cage radii are nearly linear in temperature in the whole 
glassy phase, which is natural since non harmonic effects have been neglected.

\item
The effective temperature $\teff=T/ m$ of the molecular liquid.
This temperature varies very little and it remains close to 
the Kauzmann temperature when $T$ spans  the whole low temperature phase, 
confirming the validity of our description of the 
 glass by means of a system of molecules remaining in the liquid phase. 
It is worth to stress that the linear behaviour of the parameter $m$ as a function
of $T$ is a feature shared by every 1RSB system to our knowledge.

\end{itemize}

The harmonic expansion makes sense only if  $M$ has no negative eigenvalues,
which is natural since it is intimately related to the vibration modes of the 
glass.
Notice that here we cannot describe activated processes,
and therefore we cannot see the tail of negative eigenvalues (with number 
decreasing as
$\exp(-C/T)$ at low temperatures), which is always present \cite{Ke}. 
It is known however
that the fraction of  negative eigenvalues of $M$ becomes negligible below
the dynamical transition temperature $T_D$ \cite{ScTa}
. So our harmonic expansion makes 
sense if 
the effective temperature $\teff$ is less than $T_D$.
\end{subsection}

\begin{subsection}{Small cage expansion}

\n
It is possible to introduce a slightly different way to compute 
the molecular liquid free-energy, in order to take into account:

\begin{itemize}

\item 
Non-harmonic terms. 

\item
Corrections to the quenched approximation.

\end{itemize}

\n
Starting from the  expansion of the potential in powers of the
relative variables
$u$, if one expands also the exponential of the corrective term, one
obtains an expansion of $Z_m$ as a power series
in $\app$ and $\amm$. This is the generalization to mixtures of the
small cage expansion scheme utilized in the pure case \cite{MePa1,MePa2}. 
This  expansion is not equivalent to computing perturbatively quartic and higher 
order corrections to the Gaussian approximation represented by
the harmonic re-summation. Indeed, in this $\alpha_\pm$ expansion
we are using a truncated version of the series in
(\ref{fharmoin}). On the other hand this direct expansion allows
to take into account the
annealed fluctuations of the matrix $M$ which were neglected
in the harmonic approximation. Therefore these two
types of approximations are complementary.
  In this paper we consider the harmonic re-summation and the small cage
approximation as distinct schemes of approximation and we
compare results obtained independently in both them. However, it is clear
that a better approximation of the replicated free-energy could be obtained by adding
corrections from the small cage approximation, treated in some systematic way, 
to the harmonic re-summation. A first attempt in this direction will 
be found in a following work\cite{CoPaVe}.

 The 
leading term of (\ref{zmquad}) in the $\app,\amm \rightarrow \infty$ limit is: 

\be
Z^{(0)}_m= \sqrt{\frac{2\:\pi \: \app}{m}}^{d \: N^+ \: (m-1)} 
\sqrt{\frac{2\:\pi \: \amm}{m}}^{d \: N^- \: (m-1)}
m^{d \: N/2} Z_{liq}(\beta \: m).
\label{ordine0}
\ee

\n
Accordingly, the zero-order free-energy is:

\bq
\beta \phi^{(0)}(\app,\amm,m,\beta) & = &  - \frac{1}{m N} 
\log Z_{m}^{(0)} = 
d_0 + a_0 \left ( c_+ \: \log 
\app + c_- \: \log \amm \right ),
\label{free0} 
\eq

\n
with

\bq
d_0 &  = & {d (1-m) \over 2 m} \log {2\pi \over m}-{d \over 2 m } \log{m} 
-{1 \over m \: N}  \log Z_{liq}(\beta \: m) \nn \\
a_0 & = &\frac{d \:(m-1)}{2}. \nonumber \\
\label{coeff0}
\eq

\n
The first order term is:

\bq
Z^{(1)}_m &  = & \frac{1}{N^+! \: N^-!}
\int \left ( \prod_{i=1}^N d^d z_i \right )
\left (\prod_{a=1}^m \: \prod_{i=1}^N d^d u^a_i \right )
 \left [ \prod_{i=1}^N \left ( m^d \delta ( \sum_{a=1}^m u^a_i) \right ) 
\right ] \cdot \nonumber \\
 & \cdot & \exp  \left ( 
- \frac{1}{4 \app} \sum_{a, b } \sum_{i \in +} (u^a_i - u^b_i)^2
-\frac{1}{4 \amm} \sum_{a, b } \sum_{i \in -} (u^a_i - u^b_i)^2
 -\beta \: m 
\sum_{i < j } V(z_i-z_j) \right ) \cdot \nonumber \\
& \cdot & \left ( 1 - \frac{\beta}{2}
\sum_{i < j } \sum_{a=1}^m \sum_{\mu,\nu}^d (u^a_{i \mu} -
u^a_{j \mu})  (u^a_{i \nu} - u^a_{j \nu})
V_{\mu \nu} (z_i - z_j) \right ) = \nonumber \\
& = & Z^{(0)}_m \: \left ( 1 - \frac{\beta}{2}
\left \langle \sum_{i < j } \sum_{a=1}^m \sum_{\mu,\nu}^d (u^a_{i \mu} -
u^a_{j \mu})(u^a_{i \nu} -
u^a_{j \nu}) V_{\mu \nu} (z_i - z_j) \right \rangle \right ),
\label{ordine1}
\eq

\n
from which we get the first-order contribution to the free-energy:

\bq
&\left. \right.& \beta \phi^{(1)}(\app,\amm,m,\beta)= 
 c_+ \: a_1^+ \: \app +
c_- \: a_1^- \: \amm,
\label{free1}
\eq

\n
where we define the first order coefficients as:

\bq
a_1^+ & = &\frac{d \: (m-1)}{2 \: m^2} 
\left [ c_+ \int d^d r\: \rho \: g^{++}(r) \sum_\mu V^{++}_{\mu\mu}(r)
+  c_- \int d^d r\: \rho \: g^{+-}(r) \sum_\mu V^{+-}_{\mu\mu}(r) \right ]
\nonumber \\
a_1^- & = &\frac{d \: (m-1)}{2 \: m^2} 
\left [ c_- \int d^d r\: \rho \: g^{--}(r) \sum_\mu V^{--}_{\mu\mu}(r)
+  c_+ \int d^d r\: \rho \: g^{-+}(r) \sum_\mu V^{-+}_{\mu\mu}(r) \right ]
\nonumber \\
\eq

\n
Up to first order, the harmonic re-summation and the small cage expansion give the same
results. Differences appear at the second order level, which is presented 
in appendix. In fact, the second order term in the harmonic re-summation is:

\be
\label{harmo2}
{(m-1) \over 4 m } \int d^d r\rho \sum_{\eps \epp}
g^{\eps \epp}(r) \sum_{\mu\nu} \((C^{\eps \epp}_{\mu\nu}(r)\))^2, 
\ee

\n
while the second order term in the small cage approximation adds two new kinds
of term (see the Appendix):

\begin{itemize}

\item
Those involving fourth derivatives of potential, 
which are anharmonic corrections,
are proportional to $(m-1)^2$, unlike any other term up to second order.
This means that they are less important near $\tk$ where $m \simeq 1$, and more
important at very low temperatures.

\item
Those expressing the fluctuations of the diagonal terms of $M$. These are
corrections to the `quenched' approximation. 

\end{itemize}

\n
The  free-energy per particle, up to second order, is then:

\bq
\beta \phi(\app,\amm,m,\beta) & = &
d_0 + \frac{a_0}{m} \left ( c_+ \log \app + c_- \log \amm \right)
+ c_+ \: {a_1^+ \app} + c_- \: {a_1^- \amm} +
\nn \\ & + & c_+ \: 
{a_2^{++} \app^2} + c_- \: {a_2^{--} \amm^2} +
c_+ \: c_- \: {a_2^{+-} \app  \amm)},
\eq

\n
where the coefficients $a_2^{\eps \epp}$ are given in the Appendix.

The free energy $\phi$ should be studied in the zero coupling limit,
that is $\app,\amm \to \infty$. This can not be done 
directly with a powers series of $\app,\amm$ truncated at a finite order.
Therefore one must first take the Legendre transform of $\phi$,
as previously discussed, getting the thermodynamic
potential $G$ as an expansion in powers of different 
cage sizes $A_{\eps}$, defined by means of (\ref{gabbie}).
Within this formulation, the free energy $\phi$ in the vanishing coupling limit 
is  obtained by looking for possible minima of $G$ with respect 
to $A^+$,$A^-$.

\noindent
The Lagrange transformed free energy is, at first order:

\bq
\beta G(A^+,A^-,m,\beta) & = & 
\gamma_0 + \frac{d \:(1-m)}{2m} \left (c_+ \: \log(A^+) +c_- \log(A^-) \right )
+ \nn \\
& + & c_+ \: \gamma_1^+ \: A^+ + c_- \: \gamma_1^- A^- \nonumber \\
\gamma_0 & = & a_0 + \frac{d (1-m)}{m}, \hspace{.2in} \gamma_1^+=a_1^+,
\hspace{.2in} \gamma_1^-=a_1^-,
\eq

\n
and the  saddle points equations read:

\bq
\frac{\partial G}{\partial A^+} & = & 0 \Rightarrow {A^+}^*= 
- \frac{d(1-m)}{m} \frac{1}{\gamma_1^+}= \frac{1}{\beta \: r_+},
\nonumber \\
\frac{\partial G}{\partial A^-} & = & 0 \Rightarrow {A^-}^*
- \frac{d(1-m)}{m} \frac{1}{\gamma_1^-}= \frac{1}{\beta \: r_-},
\nn \\
r_+ & = & 
c_+ \: \int d^d r\: \rho \: g_{liq}^{++}(r) \:  \frac{1}{d}
\Delta V^{++}(r) +
c_- \: \int d^d r\: \rho \: g_{liq}^{+-}(r) \:  \frac{1}{d}
\Delta V^{+-}(r),  \nn \\
r_- & = & 
c_- \: \int d^d r\: \rho \: g_{liq}^{--}(r) \:  \frac{1}{d}
\Delta V^{--}(r) +
c_+ \: \int d^d r\: \rho \: \:g_{liq}^{+-}(r) \:  \frac{1}{d}
\Delta V^{+-}(r).  
\eq

\n
The first order free energy in the
vanishing coupling limit is correspondingly given by
\bq
\beta G({A^+}^*,{A^-}^*,m,\beta) & = & 
\frac{d(1-m)}{2 m} \log \left ( \frac{2 \pi}{m} \right )
-\frac{d}{2 m} \log(m) - \frac{1}{m \: N} \log Z_{liq}(\beta m) + \nonumber \\
& - & \frac{d \:(1-m)}{2m} \left (c_+ \: \log(\beta \: r_+) +c_- \log(\beta \: 
r_-) \right ).
\label{gpom}
\eq

\n
This expression for $G$ looks quite reasonable. First of all one may note 
that in the $m \rightarrow 1$
limit it reproduces the `liquid' free energy density $\beta f=
-\log Z_{liq}(\beta)/N$,
 as it should. Moreover, in the limit in which the "$+$" and 
 "$-$" particles 
are no more distinguishable, this expression for 
$G$ coincides with the one found in the pure case \cite{MePa1,MePa2}
(More precisely, the two generalized free energy would coincide in this limit 
if the liquid free energies at inverse temperature 
$\beta \: m$ were the same, which would be true if one
would forget about the mixture entropy contribution  
$\propto c_+ \log c_+ +c_-\log c_-$).

The computation of the second order terms can be carried out in a very 
similar way. One gets (see the Appendix): 

\bq
\beta G(A^+,A^-,m,\beta) & = & \gamma_0 
 + \gamma_3 \left (c_+ \: \log(A^+) +c_- \log(A^-) \right ) + \nn \\
& +& c_+ \: \gamma_1^+ \: A^+ + c_- \: \gamma_1^- \: A^- +\nonumber \\
& + & c_+ \: \gamma_2^{++} \: (A^+)^2 + c_- \: \gamma_2^{--} \: (A^-)^2 +
c_+ c_- \gamma_2^{+-} A^+ \: A^-,
\eq
with
\be
\gamma_3 \equiv \frac{d \: (1-m)}{2m}.
\ee

In evaluating the formulae of the appendix one needs to know the three particles 
correlation 
function.  This correlation function can be computed starting from a generalized 
HNC expansion \cite{MePa1}.  Here we follow the simpler route of evaluating 
the three point 
function using the superposition principle, i.e.  
$g_{3}(x,y,z)=g(x-z)g(x-y)g(y-z)$.
\n
When looking for the minimum $\partial G/\partial A^+=0, \partial G/ 
\partial A^-=0$, one faces the problem that the second order corrections
are very important (this happens also in the pure case). In this case
the solution can be found only through a
  perturbation around the first order solution. 
In this way one gets
\bq
{A^+}^* & = & {A^+_1}^*+\delta \: {A^+_2}^* \nn \\
{A^-}^* & = & {A^-_1}^*+\delta \: {A^-_2}^* \nn \\
G({A^+}^*,{A^-}^*,m,\beta) & = & G_1 \: + \: \delta \: G_2
\eq
and, by writing $m=m_1+ \delta \: m_2$, the stationarity
condition reads:
\beq \nn
{\partial G_1 \over \partial m}(m_1)&=&0 \\
m_2&=&- {\partial G_2 \over \partial m}(m_1) \(( {\partial^2 G_1 
\over \partial m^2}(m_1)\))^{-1}.
\label{mexp}
\eq
Therefore, one looks for the value $m_1^*$ which maximizes $G_1$, which
is nothing but the first order free energy, and then one computes
 the second order corrections
at $m=m_1^*$. The result is:
\bq
{A^+_2}^* & = &  
\frac{2 \: \gamma_2^{++}}{(\gamma_1^+)^3} + c_- \frac{\gamma_2^{+-}}
{(\gamma_1^+)^2 \: \gamma_1^-}, \nonumber \\
{A^-_2}^* & = & 
\frac{2 \: \gamma_2^{--}}{(\gamma_1^-)^3} + c_+ \frac{\gamma_2^{+-}}
{(\gamma_1^-)^2 \: \gamma_1^+}, \nn \\
G_2 & = & c_+ \left ( \frac{\gamma_3}{\gamma_1^+} \right )^2 \: \gamma_2^{++}
+ c_- \left ( \frac{\gamma_3}{\gamma_1^-} \right )^2 \: \gamma_2^{--}
+ c_+ \: c_- \frac{\gamma_3^2}{\gamma_1^+ \: \gamma_1^-} \: \gamma_2^{+-}.
\eq
Finally, the second order correction to $m_1$ is obtained by (\ref{mexp}).

\end{subsection}

\end{section}

\begin{section}{HNC for binary mixtures}

In evaluating the liquid free energy $f_{liq}$ and the 
$g^{\eps \epp}$ at
the effective temperature $T/m$ we use the so-called Hypernetted Chain 
approximation (HNC), a simple closure approximation  that
consists in neglecting the `bridge' diagrams in the Mayer expansion
\cite{Han1,MoHi,HNC,Hi}.
For homogeneous fluids, apart from the constant $s_{misc}=-c_+\log(c_+)
-c_-\log(c_-)$, the free energy of the liquid in the HNC approximation is given by:

\bq
\frac{1}{N} \beta F[\{ g^{\eps \epp}(r) \}] & = &  \log \rho -1+ 
\frac{\rho}{2}\int d^d r
\sum_{\eps,\epp} c_{\eps}  c_{\epp}
\left\{ g^{\eps \epp}
(r) \left [ \log g^{\eps \epp}(r) + 
\beta V^{\eps \epp}(r)-1 
\right ] +1 \right\} + 
\nonumber \\
& - &  \frac{1}{2 \rho}  
\int \frac{d^d k}{(2 \pi)^d}\left\{\log {\cal D} - 
 \sum_{\eps} \rho \: c_{\eps} h^{\eps \eps}(k) 
+  \sum_{\eps,\epp}c_{\eps}  c_{\epp} \frac{(\rho \:  
h^{\eps \epp}(k))^2}{2} \right \}
\eq

\n
where 

\be
h^{\eps \epp}(r) = g^{\eps \epp}(r)-1,
\ee

\n
and ${\cal D}$ is the determinant of the matrix

\be
\left (
\begin{array}{cc}
1+\rho c_+ \: h^{++}(k) & \rho c_+ h^{+-}(k) \\
\rho c_- \: h^{+-}(k) & 1+\rho c_- h^{--}(k) 
\end{array}
\right ).
\ee

\n
The  closed set of HNC
equations for the two point correlations can be derived
as a stationarity condition of the functional $F$ with respect 
to these  correlation: they are 
solved using the same numerical technique utilized in the pure
case \cite{MePa2}.

HNC is expected to be a good starting point for our study since both $f$ and 
the mean values of quantities that involve only two particles correlation 
functions seem to be evaluated with an error smaller than $10 \%$ in the 
temperature region we are interested in, as we verified by comparing the 
analytical estimations with simulation results.
The terms involving the three point correlation
functions, when evaluated by the superposition approximation
and the HNC pair correlations, are reproduced with errors
which seem to be smaller than $30 \%$.

\end{section}

\begin{section}{Results and discussion}

\n
Before discussing the analytical and numerical results on the 
soft sphere binary mixture, let us pay  attention, for a while, to 
the soft sphere model (\ref{pot}) with the particular value $R = 1$ (i.e.,
the pure case). 
This allows to compare thermodynamic quantities obtained 
within the small cage expansion up to second order (evaluating the 
three point function by the superposition approximation)
with those computed at the same order in the replicated HNC re-summation 
scheme \cite{MePa1}.

\begin{figure}[htbp]
\begin{center}
\leavevmode
\epsfig{figure=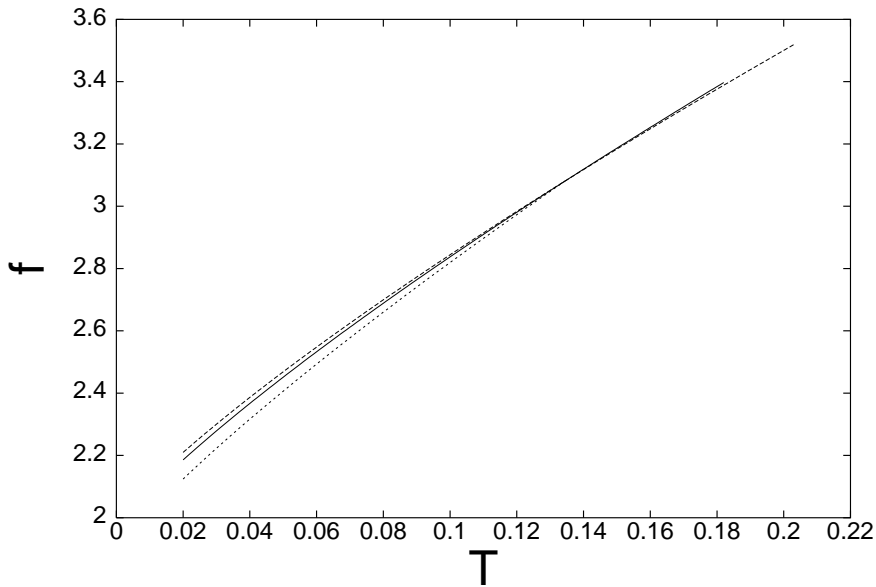,angle=270,width=11cm}
\vspace{.5cm}
\caption{The free energy of the pure soft sphere model
versus temperature. The three curves are the results obtained
from the small cage expansion at the first order (dotted line)
and at the second order (dashed line),
and those from the HNC re-summation scheme (continuous line).}
\end{center}
\label{r1a}
\end{figure}

\begin{figure}[htbp]
\begin{center}
\leavevmode
\epsfig{figure=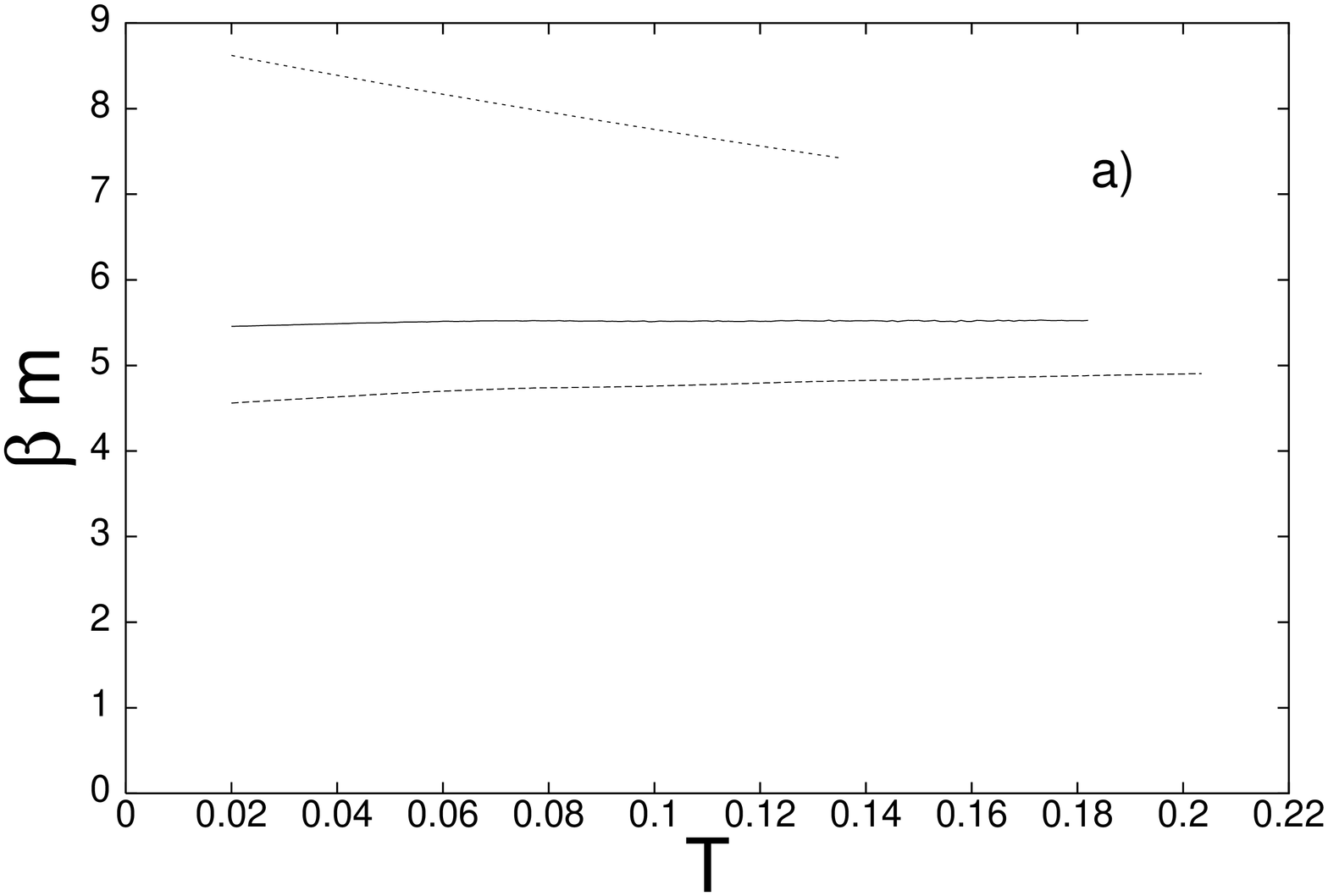,angle=270,width=8cm}
\epsfig{figure=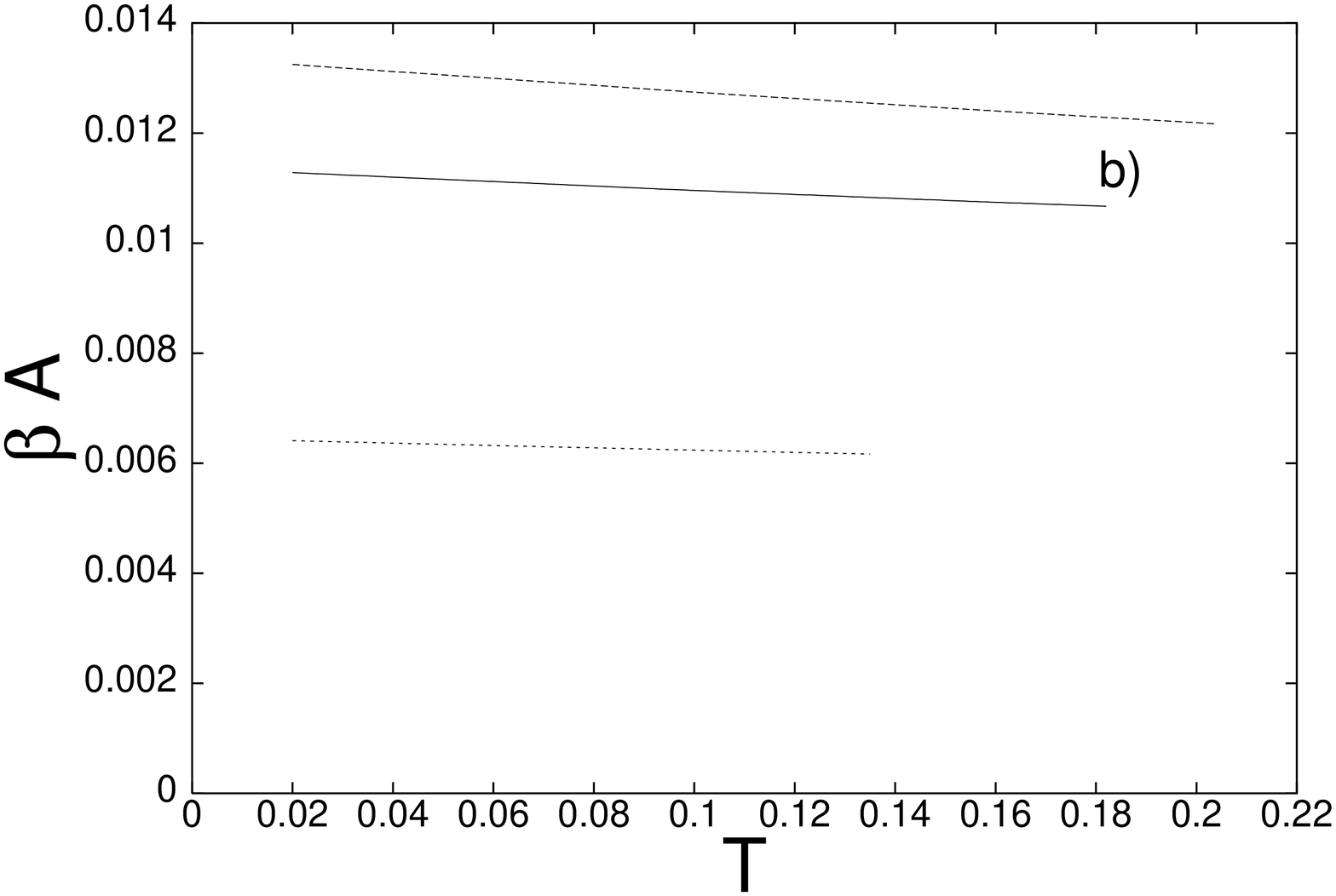,angle=270,width=8cm}
\vspace{.5cm}
\caption{$\beta \: m$ ($a$) and $\beta A$ ($b$) 
of the pure soft spheres model
versus temperature. The three curves are the results obtained
from the small cage expansion at the first order (dotted line)
and at the second order (dashed line),
and those from the HNC re-summation scheme (continuous line).
}
\end{center}
\label{r1b}
\end{figure}

In [Figg. 1-2a,b] 
we show the free energy, the effective
inverse temperature $\beta m$ and the cage radius $A$ for the 
pure soft sphere model both at the first order, that  
gives the same results in the two cases, and at the second one. 
As already outlined, when 
starting from the generalized HNC expression, the second order coefficient 
$\gamma_2$ is obtained without further approximations than the one related 
to the use of HNC. On the other hand, these results confirm that evaluating 
the three point correlation function which appears in $\gamma_2$ by the 
superposition approximation is a rather good approximation. 
In particular we get very similar values for the thermodynamic
transition point, $\Gamma_K \simeq 1.53$ from the HNC resummation scheme
and $\Gamma_K \simeq 1.49$ when using small cage expansion, i.e.
an error less than  $3\%$.

Now we come back to
the soft spheres binary mixture with the interaction parameters described in 
(\ref{pot}), 
taking in particular the value $R=1.2$ of the ratio between the effective 
diameters in order to
obtain analytical results comparable to the numerical ones. We consider both 
the small cage expansion to 
second order and the harmonic re-summation, finding results in very good 
agreement as  is shown 
in [Fig. 3], where the glassy phase free energy computed in the two 
different schemes of 
approximation is plotted as a function of $T$ ( for simplicity we take in the 
following $\rho=1$).

\begin{figure}[htpb]
\centerline{\epsfig{figure=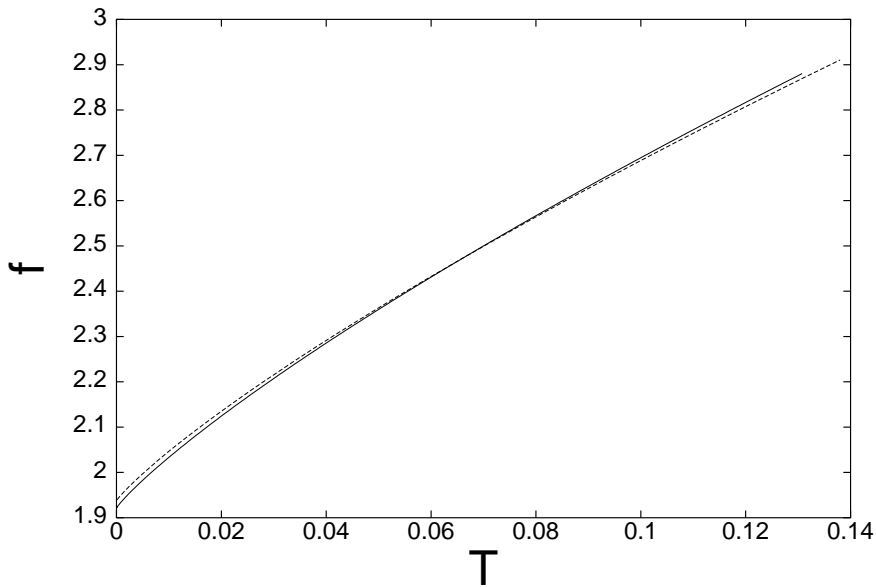,angle=270,width=11cm}}
\vspace{.5cm}
\caption{Free energy of the soft sphere mixture vs temperature.
The continuous
line  is the result of the 
 harmonic resummation scheme  and the dashed line is the result of the
small cage expansion to second order.}
\label{enlibera}
\end{figure}
The evaluations of the thermodynamic critical temperature obtained by the two 
analytic methods 
 nearly coincide: we get $\Gamma_K \equiv T_K^{-1/4} \simeq 1.65$, which is 
in agreement with the 
numerical estimates that we are going to discuss.  For the sake of comparison, 
let us remember that the
Mode Coupling critical value for this model \cite{Han2}
is $\Gamma_D \simeq 1.45$.  
Let us note that the 
ratio $T_D/T_K$ is usually found to be between 1.2 and 1.6. 
  
We stress that the parameter $m$ and cages size, $A^+$ and $A^-$,  
plotted in 
[Figg. 4a,b] are nearly linear with temperature. 
This means, in particular, that the effective temperature $T/ m$ is always 
close to $T_K$, so in our theoretical computation we need
only the mean values of observables in the liquid phase, at temperatures 
where the HNC approximation still works quite well. 

One can also observe that the specific heat (see [Fig. 5b]) shows an 
evident `jump' 
at $T_K$, remaining close to the crystal-like value, $3/2$ (we have not
included the kinetic energy), 
 in  the whole glassy phase. The 
qualitative behavior of thermodynamic quantities, apart from the presence of 
the two distinct radii, 
is very similar to that observed in the pure case \cite{MePa1,MePa2} and
it corresponds to a second order transition from the thermodynamic
point 
of view.

\begin{figure}[htbp]
\begin{center}
\leavevmode
\epsfig{figure=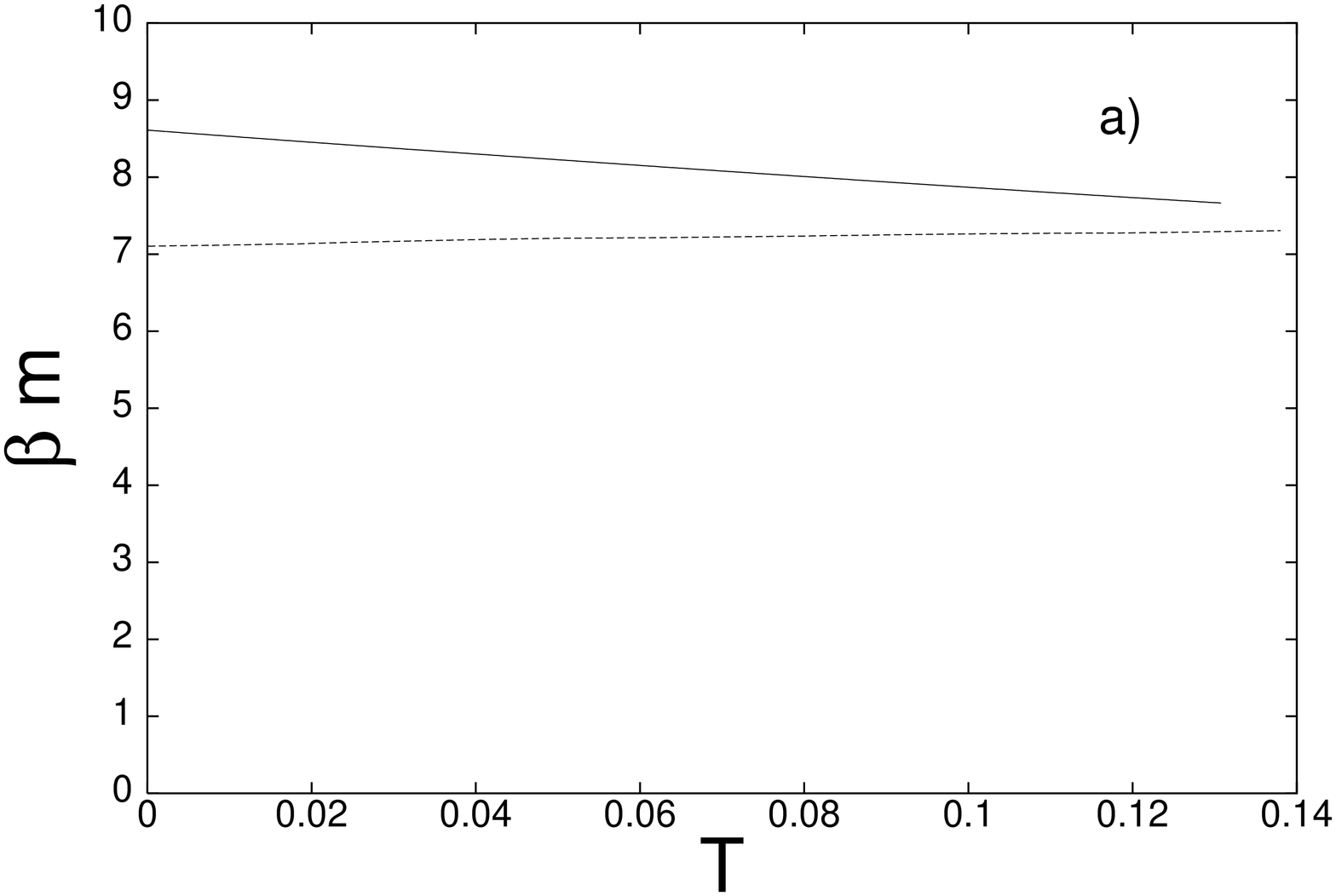,angle=270,width=8cm}
\epsfig{figure=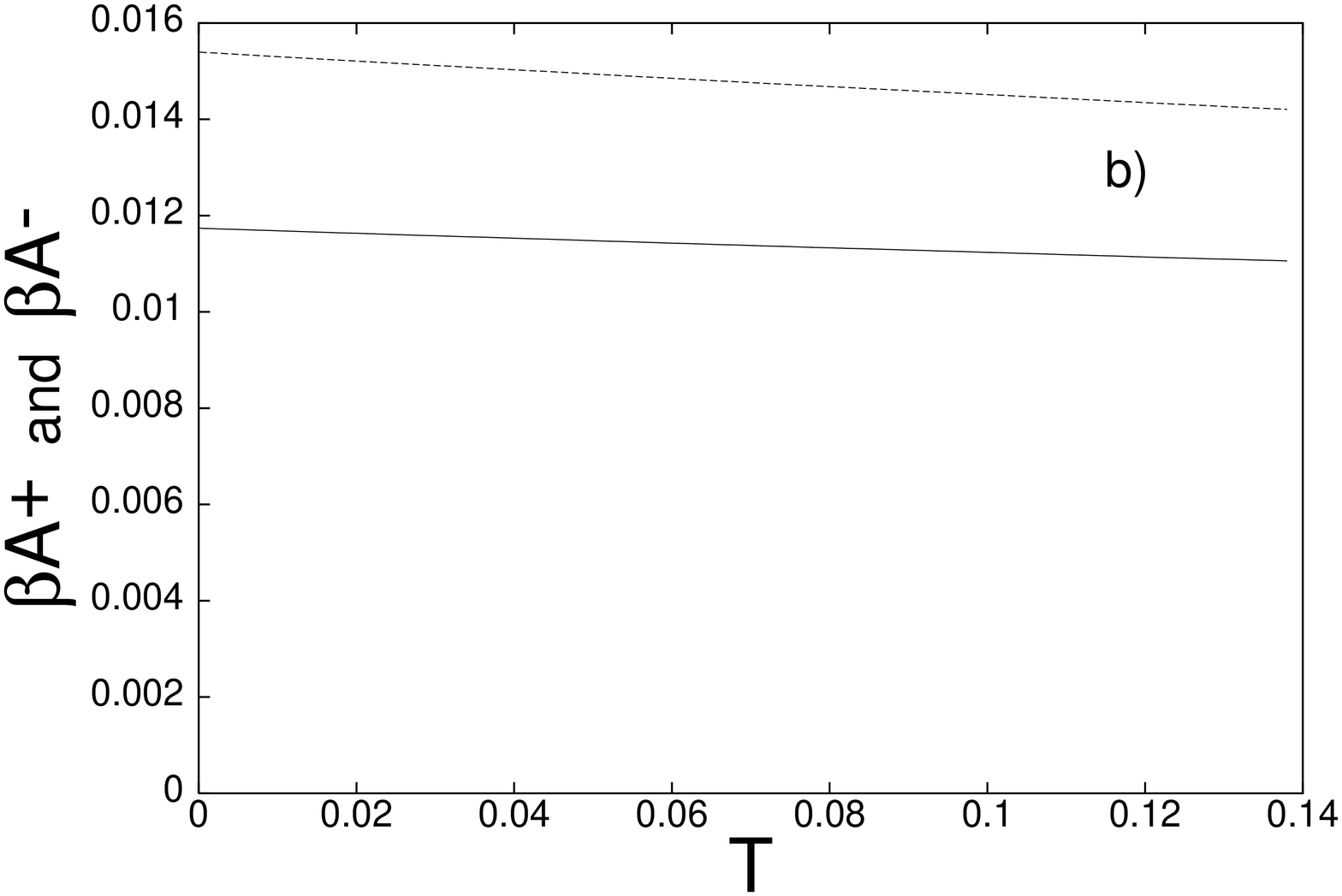,angle=270,width=8cm}
\vspace{.5cm}
\caption{In ($a$) we plot $\beta \: m$ vs temperature, 
from  the harmonic re-summation scheme (continuous line) 
and from the low temperature expansion to second order (dashed line).  
In ($b$) we 
present $\beta A^+$ (continuous line) and $\beta A^-$ (dashed line) 
computed in the low temperature expansion to second order. Note that, quite 
reasonably, the smallest
cage radius corresponds to particles with the largest effective diameter.}
\end{center}
\label{emme}
\end{figure}

\begin{figure}[htbp]
\begin{center}
\leavevmode
\epsfig{figure=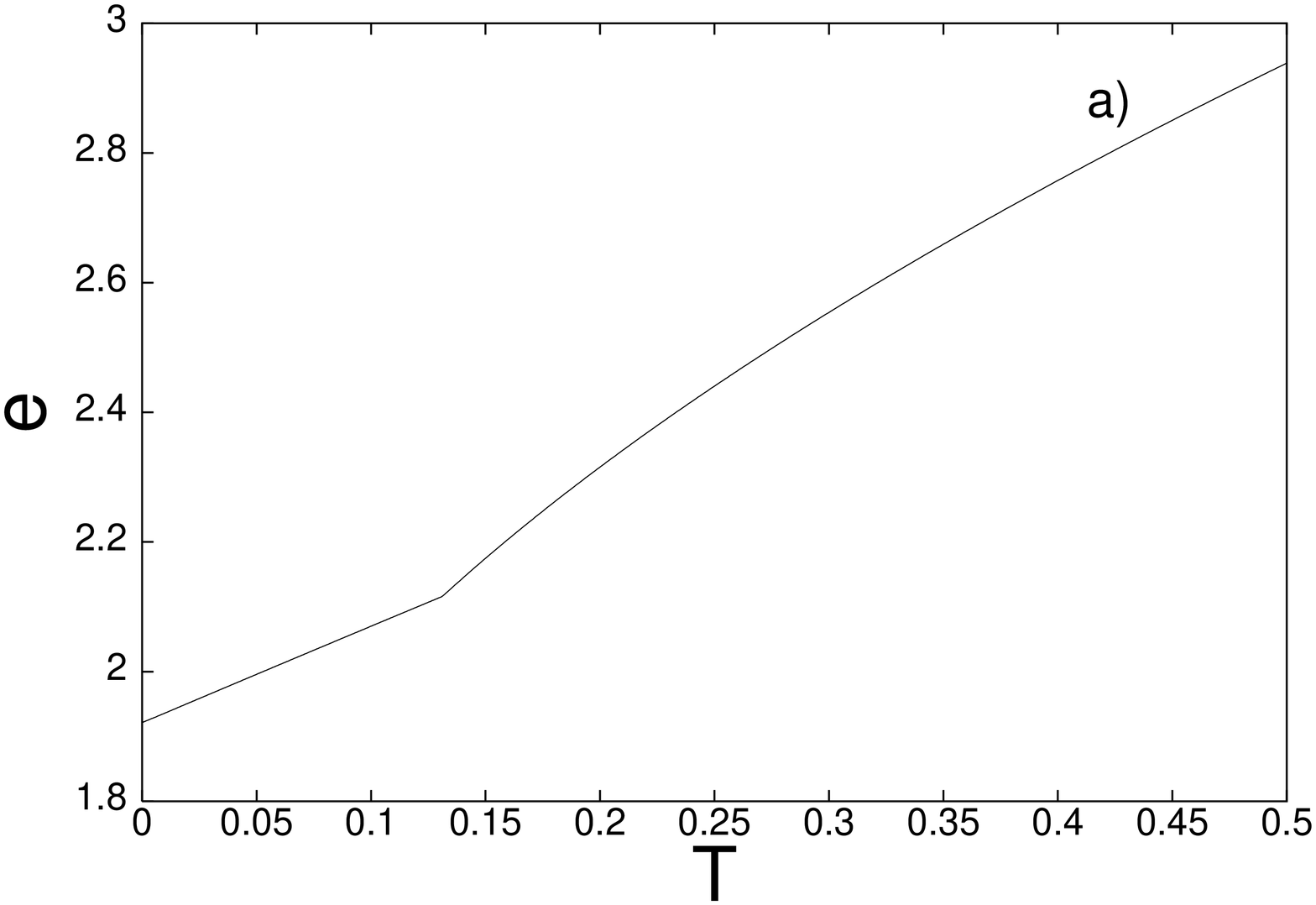,angle=270,width=8cm}
\epsfig{figure=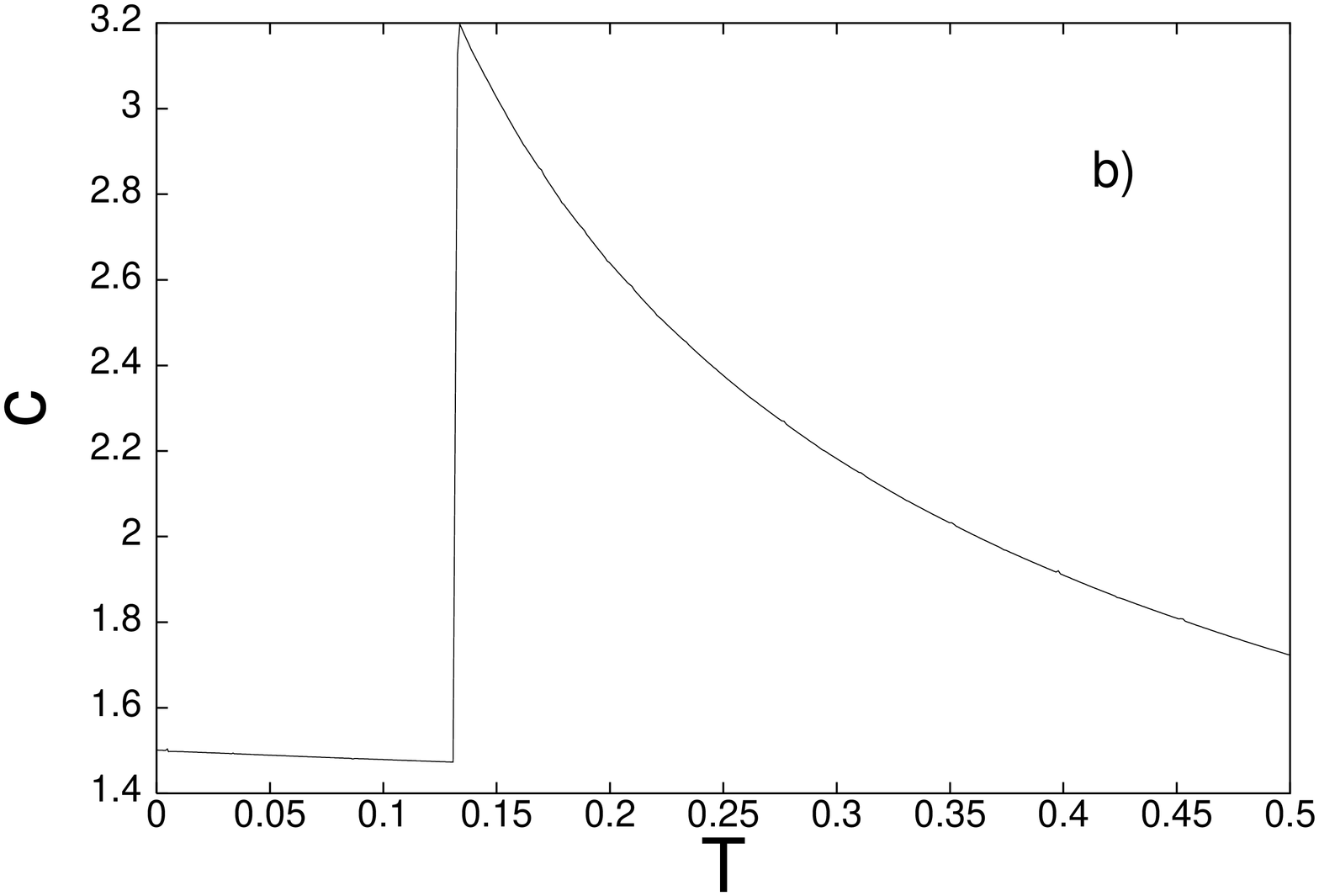,angle=270,width=8cm}
\vspace{.5cm}
\caption{The energy ($a$) and the specific heat ($b$)
of the soft sphere mixture versus temperature, 
both in the liquid and in the glassy phase, from the harmonic re-summation 
scheme.}
\end{center}
\label{enec}
\end{figure}

The harmonic re-summation scheme suggests an intriguing approach for
evaluating the thermodynamic critical temperature by simulations, 
starting from (\ref{comple}). Here the liquid entropy can be obtained 
for instance by numerically integrating the energy
\be
S_{liq}(\beta)=\beta \left ( E_{liq}(\beta)-F_{liq}(\beta) \right )=
S_{liq}^0 + \beta \: E_{liq}(\beta) -\int_0^{\beta} d \beta' E_{liq}(\beta')
\ee
where $S_{liq}^0$ is the entropy of the perfect gas in the 
$\beta \rightarrow 0$ limit, i.e in the binary mixture case
\be
S_{liq}^0=N \: \left ( 1-\log \rho-c_+\log c_+ -c_-\log c_- \right).
\ee
Moreover, one can think of directly numerically evaluating the `harmonic solid' 
entropy 
\be
{S_{sol}(\beta) \over N}= {d \over 2} (1+\log(2 \pi)) - {1\over 2 \: N}
\left \langle  \Tr \log (\beta {\cal M} ) \right \rangle, 
\ee
by diagonalizing the `instantaneous' Hessian and by averaging over different 
configurations.
The knowledge of $S_{liq}$ and $S_{sol}$ allows  to obtain a numerical 
estimate of $\tk$ as the 
temperature where the two entropies become equal, and to measure the complexity
\be
\Sigma(\beta)=\frac{1}{N} \left [ S_{liq} (\beta) - S_{sol}(\beta) \right ].
\ee

When attempting to obtain such evaluations, we face two kind of problems:
\begin{itemize}
\item The well known hard task of thermalizing glass-forming liquids at low 
temperatures. 
Here we choose to perform a simulated annealing run of a quite large system, 
using data on the liquid energy down to the temperature where the equilibrium 
was still 
reachable in a reasonable CPU time ($\Gamma \sim 1.5$). 
Then we extrapolate the liquid entropy behavior at lower temperatures by 
fitting data
in the interval $\Gamma \in [1,1.5]$ with the power law
\be
S_{liq}(T)=a \: T^{-2/5}+b.
\ee

In fact, it has been shown \cite{RoTa} that the potential energy of
simple liquids at high densities and low temperature must follow this law,
and we find that our numerical data are in a very good agreement with it.

\item The correct evaluation of the solid entropy, which is
a  subtle task.
Beyond the mean field approximation there always exists a non zero 
number of 
negative eigenvalues, which decreases as $\exp(-C/T)$ at low temperatures 
\cite{Ke} 
and is expected to be negligible below the Mode Coupling 
temperature. 
An estimate of the error on $S_{sol}$ can be found by doing the following
two measurements. a) One includes in the computation of
$\Tr \log (\beta {\cal M} )$ only the ${\cal N}_{pos}$ positive eigenvalues. b)
One includes all eigenvalues, but one takes the absolute 
values of 
the negative ones: 
\bq
{S^{(a)}_{sol} \over N} ={d \over 2} \left [ (1+\log({2 \pi \over \beta})) - 
\langle \frac{1}{{\cal N}_{pos}} \sum_{i=1}^{{\cal N}_{pos}} \log {\lambda_i} 
\rangle
\right ] \\
{S^{(b)}_{sol} \over N} ={d \over 2} \left [ (1+\log({2 \pi \over \beta})) - 
\langle \frac{1}{{ d N}} \sum_{i=1}^{d N} \log {|\lambda_i|}
\rangle \right ]. 
\eq
The percentage of non positive eigenvalues that we find by diagonalizing the 
`instantaneous' Hessian
is still about $20 \%$ at $\Gamma \sim 1$, it decreases to less than $10 \%$ at 
$\Gamma \sim 1.2$
and in the region definitely below $T_D$, i.e. above $\Gamma \sim 1.5$, it is 
$\sim  4 \%$. 
On the other hand, nearly all the negative eigenvalues are less than one in
absolute value. Therefore, particularly at temperatures $T \simg T_D$, we find a 
sizable
difference between $S^{(a)}_{sol}$ and $S^{(b)}_{sol}$ (we disregard in both 
cases the very few 
$|\lambda| < 10^{-4}$), as  is shown in [Fig. 6]. One should note 
that
$S^{(b)}_{sol}$ seems to display the most regular behavior as a function of 
$\Gamma$. 
The presence of negative eigenvalues is possibly related
also to the fact that when diagonalizing the `instantaneous' Hessian the system 
can be 
far from the `center' of the minimum, in positions where higher order 
corrections to a
harmonic approximation of the energy landscape are important.

A  more extensive study should be performed in order to better understand 
these subtleties of the computation of the solid entropy. 
However we would like to mention here a third way for 
evaluating  numerically the solid entropy. Starting from an equilibrium configuration at a 
given 
$\Gamma$ value, we performed a Monte Carlo run at $T=0$, allowing only quite 
small
displacements to each particle. The percentage of non positive eigenvalues 
becomes very rapidly 
$< 2 \%$  in the whole temperature range considered and correspondingly the 
two
different ways of evaluating the `solid entropy' give compatible results. The 
obtained
$S_{sol}^{(c)}$ is near to the one evaluated from the `instantaneous' Hessian by 
using
also the absolute values of negative eigenvalues in the region $T \sim T_D$ 
but it decreases slightly faster
when lowering the temperature (see [Fig. 6]).
 \end{itemize}
 
More precisely, we performed a simulated annealing run of a system of $N=258$ 
particles,
in a cubic box with periodic boundary conditions, starting from $\Gamma=0.05$ 
and performing
up to $2^{22}$ MC steps at each $\Delta \Gamma=0.05$, the maximum shift 
$\delta_{max}$ permitted to each
particle in one step being chosen such as to get an acceptance rate
$ \sim 0.5$. 
The energy
and its fluctuation were measured in the last half of the run at a
given $\Gamma$-value.

Just for decreasing the error on the evaluation of $S_{liq}$, we fit the very 
high temperature
data on the energy, up to $\Gamma_0=0.2$, by using $\Gamma^{3} E(\Gamma)= a 
\Gamma^2+b \Gamma+c$, 
obtaining correspondingly $F(\Gamma_0)=4 a \Gamma_0^3/3+2b\Gamma_0^2+4c\Gamma_0$ 
that turns out
to be perfectly compatible with the HNC value (i.e. we are still in the region 
where no differences 
are observable between numerical data and the HNC approximation). 
The integration is 
subsequently performed by interpolating with a standard numerical subroutine 
the simulation data in order to get a result independent on the integration 
interval.

In order to evaluate $S^{(a)}_{sol}$ and $S^{(b)}_{sol}$ we considered 16 
different configurations
in the last half of the run at each $\Gamma$-value, while $S^{(c)}_{sol}$ was 
measured from
the configurations obtained by these ones with 5000 MC steps at $T=0$ (starting 
from
 $\delta_{max}=0.1$ and decreasing it up to 0.02 during the run). One should
note that at a given $\Gamma$-value the obtained evaluations of $S^{(c)}_{sol}$ 
seem to depend weakly on the starting equilibrium configuration (i.e. 
fluctuations are 
very small). Moreover we get perfectly compatible results both halving and 
doubling the number 
of MC steps (in the last case we find practically only positive eigenvalues).

\begin{figure}[htpb]
\centerline{\epsfig{figure=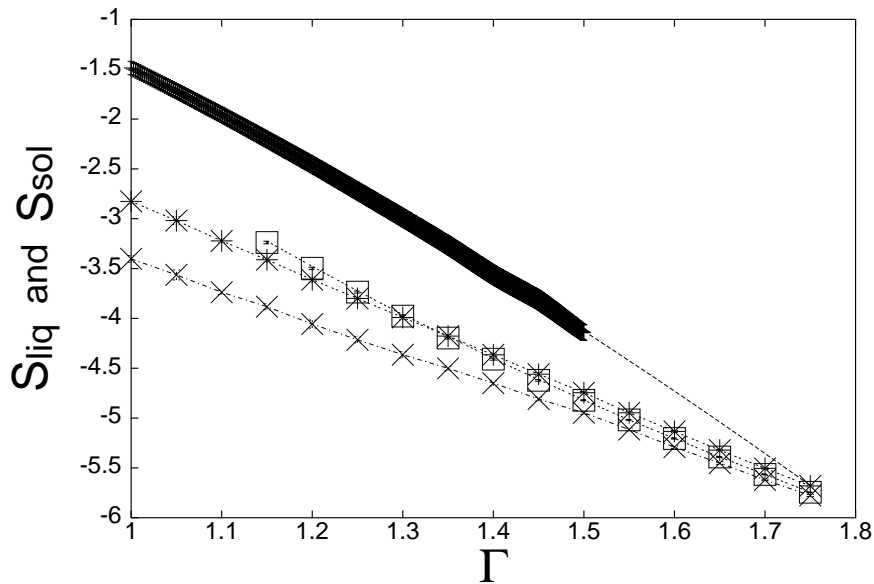,angle=270,width=11cm}}
\vspace{.5cm}
\caption{The entropy of the liquid (+) and the different evaluations (see 
text) 
of the amorphous solid entropy, $s^{(a)}_{sol}$ ($\times$), $s^{(b)}_{sol}$ 
($\ast$) and 
$s^{(c)}_{sol}$ ($\Box$), as functions of $\Gamma$,
obtained in the numerical simulation. In the liquid entropy case 
the line is the
best fit to the power law $s_{liq}=a \beta^{2/5}+b$, otherwise lines are only 
interpolations between neighboring points.
The liquid and solid entropies seem to cross around $\Gamma_K \sim 1.75$, 
which is the corresponding
estimation of the thermodynamic liquid-glass transition.}
\label{entropie}
\end{figure}

We plot in [Fig. 6] both $s_{liq}(\Gamma)$ and the obtained 
evaluations of 
$s_{sol}(\Gamma)$ by the different ways considered. $s^{(a)}_{sol}$, 
$s^{(b)}_{sol}$ and  
$s^{(c)}_{sol}$ are very close to each other
when approaching the thermodynamic liquid-glass transition, giving similar 
estimates of
$\Gamma_K \sim 1.75$.

The study of the system coupled to a reference configuration $x_{ref}$, which is 
an equilibrium
configuration of the system itself at the considered temperature, allows to 
measure the complexity by an 
alternative route. One considers
\be
\beta {\cal H}=\beta {\cal H}_0+\epsilon (x-x_{ref})^2, 
\ee
where
\be
(x-x_{ref})^2 \equiv \sum_{i=1}^N \sum_{\mu=1}^d (x_i^{\mu}-
{x^{\mu}_{ref}}_i)^2
\ee
is the squared distance between the configurations (note that the coupling 
breaks the 
rotation-translation-permutation invariance). Therefore
\be
\beta f(\epsilon,\beta)=\beta f_0(\beta) 
+ \int_0^{\epsilon} d\epsilon' \langle  (x-x_{ref})^2 \rangle_{\epsilon'}, 
\ee
where in the region $T \siml T_D$ 
\be
\beta f_0(\beta)=\lim_{\epsilon \rightarrow 0^+}f(\epsilon,\beta)\simeq
\beta e(\beta)-\Sigma(\beta) 
\ee
On the other hand, one has
\be
\lim_{\epsilon \rightarrow \infty} \beta f(\epsilon,\beta)= \beta 
f_{\infty}(\beta)
=\beta e(\beta)+\frac{d}{2} \left ( \log  (
\frac{ \epsilon}{2\pi} )-1 \right ).
\ee
This means that one can obtain an evaluation of the configurational entropy as
\be
\Sigma(\beta) \simeq s_{liq}^0 +  \int_0^{\epsilon} d\epsilon' \langle  
(x-x_{ref})^2 
\rangle_{\epsilon'}-\frac{d}{2} \left ( \log  (
\frac{\epsilon}{2\pi} )-1 \right ),
\ee
in the large $\epsilon$ limit, taking into account as usual
the perfect gas binary mixture entropy.

Here we considered a large system of $N=2000$ particles in a cubic box with 
periodic boundary conditions 
and we put
a cut-off on the potentials, i.e. $\tilde{V}^{\epsilon \epsilon'}(r)=V^{\epsilon 
\epsilon'}(R_{max})$ 
for $r > R_{max}$, choosing
$R_{max}=1.7$ that means a practically negligible $V^{\epsilon 
\epsilon'}(R_{max}) \sim 10^{-3}$. 
The algorithm is then 
implemented in such a way that for each particle the map of the ones which are 
at distance lower than 
$R_{max}+ 2 \delta_{max}$ is recorded and updated during the run. 
 
We performed up to ${\cal N}=2^{21}$ MC steps at each considered $\Gamma=1.4$, 
1.6, 1.8, 2.0. At the end 
of the run, the configuration was copied in the reference one and subsequently a 
run of ${\cal N}/16$ MC 
steps was performed on the coupled system for different $\epsilon$ values, $ 
\epsilon=1,$ 2, 4, 8, 
$\dots$ up to very large $\epsilon \sim 10^4$, measuring the squared distance. 
We note that perfectly compatible results were obtained for ${\cal N}=2^{19}$.
The integrals were evaluated by interpolating with a standard numerical 
subroutine between the 
simulation data in order to obtain
results independent on the integration interval.

\begin{figure}[htbp]
\begin{center}
\leavevmode
\epsfig{figure=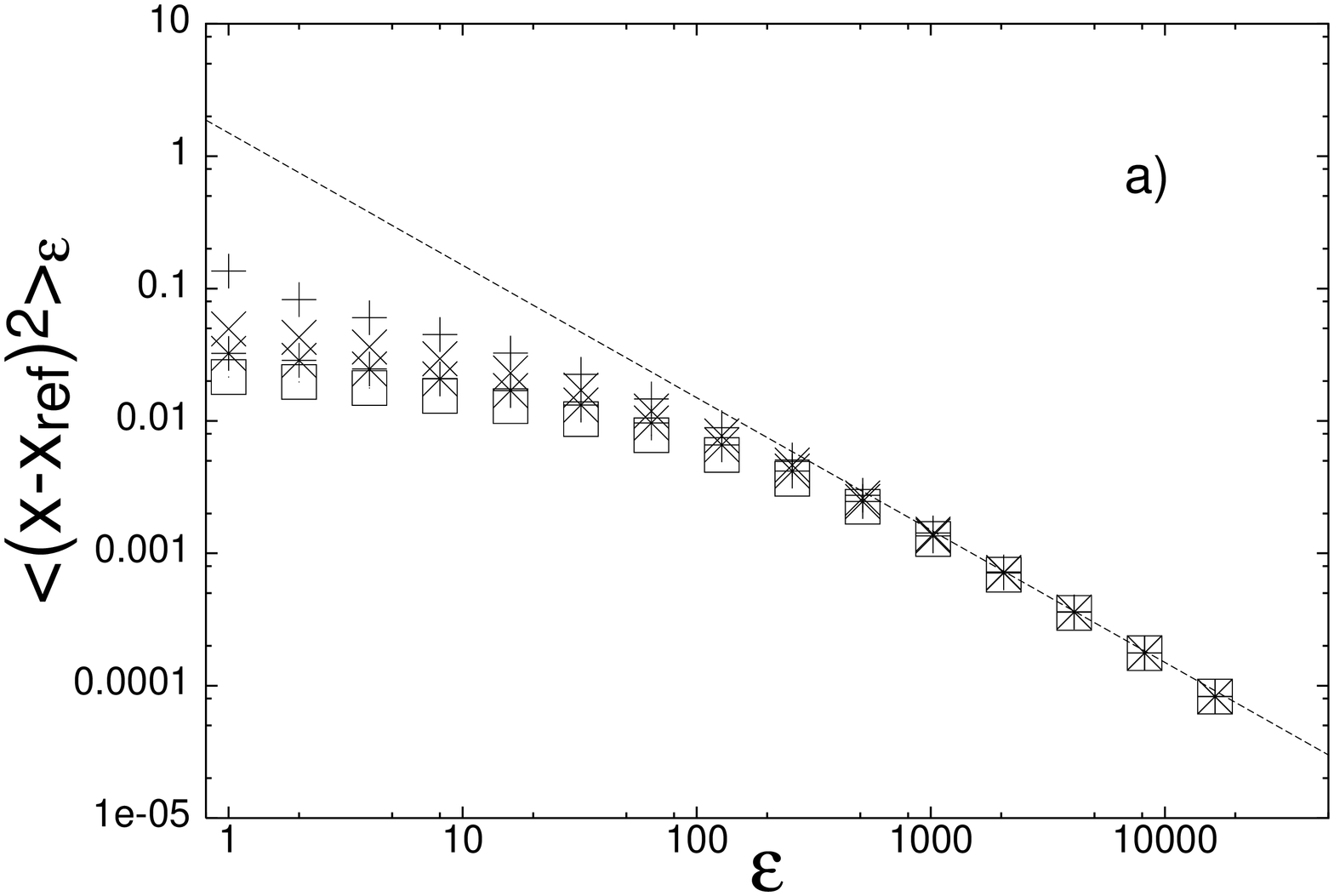,angle=270,width=8cm}
\epsfig{figure=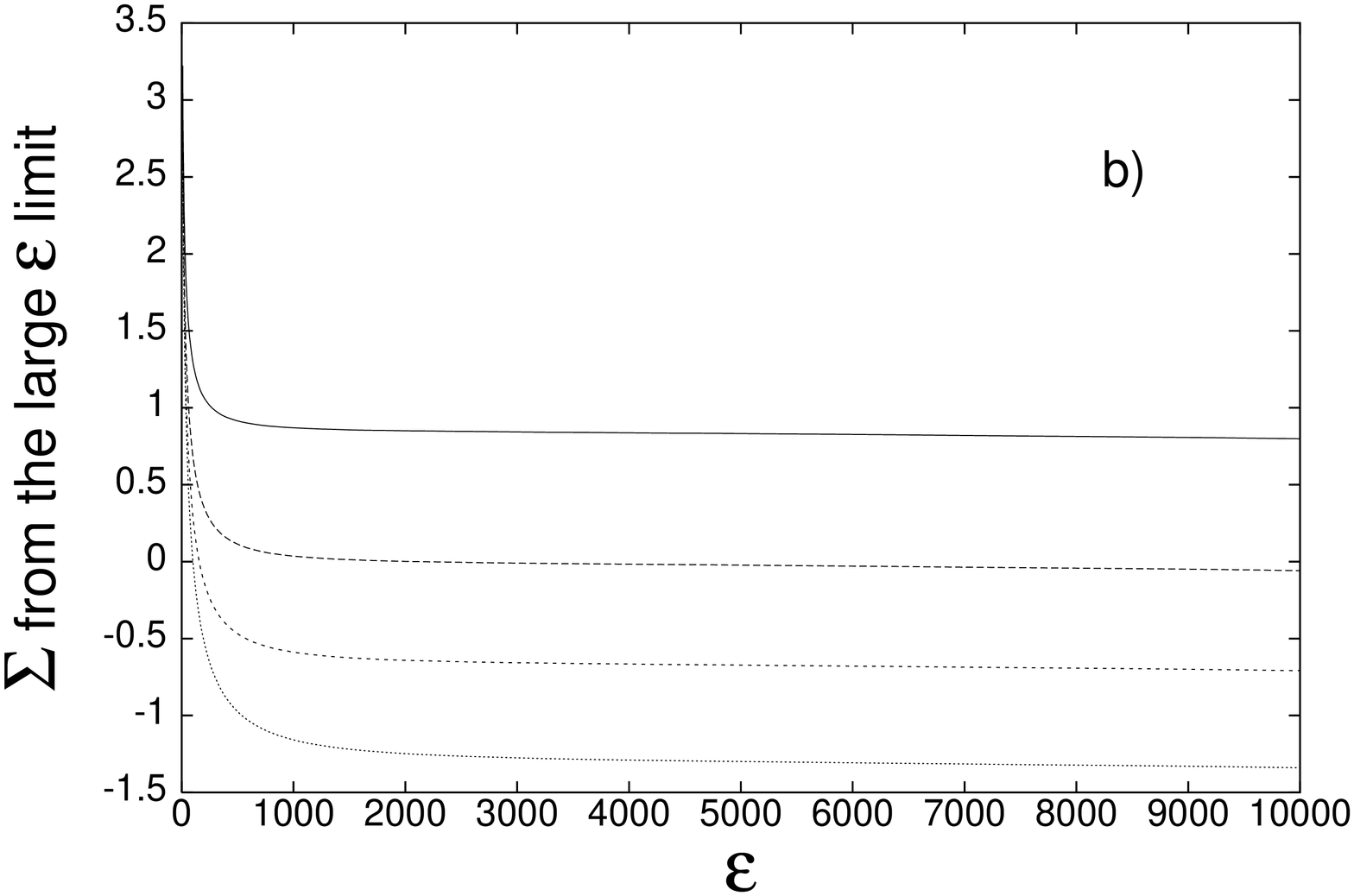,angle=270,width=8cm}
\vspace{.5cm}
\caption{In ($a$) we plot $\langle  (x-x_{ref})^2 \rangle_\epsilon$ as a 
function of $\epsilon$ at
the considered values $\Gamma=1.4$ (+), 1.6 ($\times$), 1.8 ($\ast$) and 2.0 
($\Box$). Here we show also
$3/(2\epsilon)$ (the dashed line) in order to make evident the reaching of the 
asymptotic behavior.
In ($b$) we present the evaluations (see text) of $\Sigma\simeq  
s_{liq}^0 +  \int_0^{\epsilon} d\epsilon' \langle  (x-x_{ref})^2 
\rangle_{\epsilon'}-3 \left ( \log  (
{\epsilon}{/(2\pi)} )-1 \right )/2$ in the large $\epsilon$ limit (the
different curves, from top to 
bottom, correspond to
$\Gamma=1.4$, 1.6, 1.8 and 2.0 respectively). The complexity turns out to be 
compatible with zero at
$\Gamma_K=1.6$ which is therefore the evaluation of the thermodynamic 
liquid-glass transition 
temperature.
}
\end{center}
\label{cmplx_eps}
\end{figure}

We plot in [Fig. 7a] both the data on $\langle  (x-x_{ref})^2 
\rangle_\epsilon$ 
at different $\Gamma$ as function of $\epsilon$ and $3/(2\epsilon)$.
The asymptotic behavior seems to be reached around $\epsilon=2000$, though also 
at larger
$\epsilon$ there are very weak deviations from it.
When looking at the difference between the corresponding integrals 
and $d ( \log  ({\epsilon}/{2\pi} )-1)-s_{liq}^0$ in the large $\epsilon$ limit 
(see [Fig. 7b]), one finds
that the complexity is compatible with zero at $\Gamma_K \sim 1.6$, a value 
slightly lower
than the previously obtained $\Gamma_K \sim 1.75$ (the analytical estimation 
being 
$\Gamma_K \simeq 1.65$). On the other hand, the `errors' on these estimations 
are difficult to evaluate
but they might be  quite large, a more extensive numerical analysis 
being necessary in order to 
improve these results.

\begin{figure}[htbp]
\begin{center}
\leavevmode
\centerline{\epsfig{figure=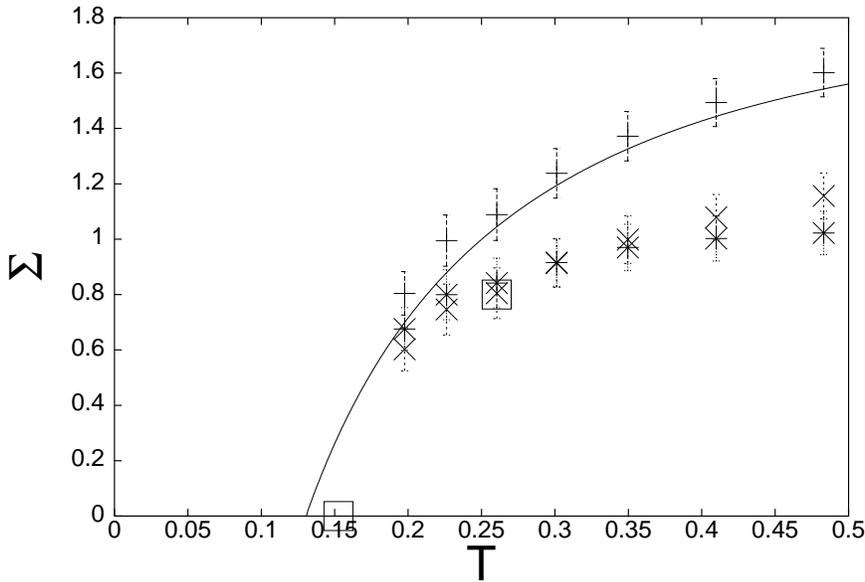,angle=270,width=11cm}} 
\vspace{.5cm}
\caption{The complexity $\Sigma (T)$ computed in the harmonic re-summation scheme 
(continuous line) 
and the different numerical evaluations, i.e. $s_{liq}-s^{(a)}_{sol}$ (+),
$s_{liq}-s^{(b)}_{sol}$ ($\times$), $s_{liq}-s^{(c)}_{sol}$ ($\ast$). The $\Box$ 
correspond to 
the $\Sigma$-values obtained by studying the coupled system at $\Gamma=1.4$ and 
1.6.}
\end{center}
\label{complex}
\end{figure}

At last we plot in [Fig. 8] the different numerical estimations of 
the 
configurational entropy $\Sigma$ as a function of the temperature and the 
behavior obtained analytically
in the harmonic re-summation scheme.
In spite of the `uncertainties' in the measures of $\Sigma$ and in the 
analytical approximations 
(first of all related to the use of the HNC closure for evaluating 
liquid quantities), the agreement between theory and simulation
looks quite satisfactory. We leave for future work both more extensive numerical 
studies 
and the improvement of the analytical results, that should allow a 
more careful comparison.

\end{section}

\section*{Acknowledgments}
The work of MM has been supported in part by the
National Science Foundation under grant No. PHY94-07194. BC would like
to thank the Physics Department of Rome University `La Sapienza`
where this work was partially developed during her PhD.


\section*{Appendix: Second order coefficients}

\n
The second order expression of replicated partition function in small
cage approximation is:

\bq
Z_{m}^{(2)} & = & Z_{m}^{0} \left \{ 1 -\frac{\beta}{4} 
\left \langle \sum_{i \neq j} \sum_{\mu \nu} V^{\al_i\beta_j}_{\mu \nu}
(z_i-z_j)
\sum_a (u_{i \mu}^a-u_{j \mu}^a)(u_{i \nu}^a-u_{j \nu}^a) \right \rangle
 \right. + \nonumber \\
& - & \frac{\beta}{2 \cdot 4!} \left
\langle \sum_{i \neq j} \sum_{\mu \nu \eta \tau} V^{\al_i\beta_j}_
{\mu \nu \eta \tau}(z_i-z_j)
\sum_a (u_{i \mu}^a-u_{j \mu}^a)(u_{i \nu}^a-u_{j \nu}^a) 
(u_{i \eta}^a-u_{j \eta}^a)(u_{i \tau}^a-u_{j \tau}^a)\right \rangle
+ \nonumber \\
& + & \left . \frac{\beta^2}{2 \cdot 16} \left \langle \left ( \sum_{i \neq j} 
\sum_{\mu \nu} V^{\ei \ej}_{\mu \nu}(z_i-z_j)
\sum_a (u_{i \mu}^a-u_{j \mu}^a)(u_{i \nu}^a-u_{j \nu}^a) \right )^2 
\right \rangle \right \},
\label{z2m}
\eq

\n
where distinction between terms which
involve sums over particles of a given kind is required. 
Taking the logarithm of partition function, one finds the
second order contribution to $\phi$

\bq
\beta \phi^{(2)}(\app,\amm,\beta) & = & 
c_+ \:
a_2^{++} \: \app^2 + c_- \: {a_2^{--} \: \amm^2} +
c_+ \: c_- \: a_2^{+-} \: \app \cdot \amm.
\eq

\n
The second order coefficients also depend on the three points
correlation functions
\be
g^{\epsilon \epsilon' \epsilon''}(r_1,r_2)= \frac{1}{c_{\epsilon}
c_{\epsilon'} c_{\epsilon''} \rho^2 N} \left
\langle \sum_{i \in \epsilon, j \in \epsilon', k \in \epsilon''} 
\delta(x_i-x_j-r_1) \delta(x_i-x_k-r_2) \right \rangle.
\ee
One has

\bq
a_2^{++} & = &{\beta \over 4} {(1-m)^2 \over m^4} 
\left [c_+ \int d^d r\: \rho \: g^{++}(r) \sum_{\mu \nu} 
V^{++}_{\mu\mu\nu\nu}(r) + \frac{c_-}{2} 
\int d^d r\: \rho \: g^{+-}(r) \sum_{\mu \nu} 
V^{+-}_{\mu\mu\nu\nu}(r) \right ] + \nn\\ 
&-& {\beta^2 \over 4} {(m-1) \over m^3} \left [ 
c_+^2 \int d^d r_1 \: d^d r_2 \: \rho^2
g^{+++}(r_1,r_2) \sum_{\mu \nu} V^{++}_{\mu\nu}(r_1) \: 
V^{++}_{\mu\nu}(r_2)+ \right. \nn \\ & + &
2 \: c_+ \: c_- \int d^d r_1 \: d^d r_2 \: \rho^2
g^{++-}(r_1,r_2) \sum_{\mu \nu} V^{++}_{\mu\nu}(r_1) \: 
V^{+-}_{\mu\nu}(r_2)+ \nn \\ & + & \left.
c_-^2 \int d^d r_1 \: d^d r_2 \: \rho^2
g^{+--}(r_1,r_2) \sum_{\mu \nu} V^{+-}_{\mu\nu}(r_1) \: 
V^{+-}_{\mu\nu}(r_2) \right ]
+ \nn \\ 
&-& {\beta^2 \over 2} {(m-1) \over m^3} 
\left [ c_+\int d^d r \: \rho \: g^{++}(r) 
\sum_{\mu \nu} V^{++}_{\mu\nu}(r) \: V^{++}_{\mu\nu}(r)+ \right. \nn \\ & + &
\left. \frac{c_-}{2}\int d^d r \: \rho \: g^{+-}(r) 
\sum_{\mu \nu} V^{+-}_{\mu\nu}(r) \: V^{+-}_{\mu\nu}(r) \right ] \nn \\
a_2^{+-}& =& {\beta \over 4} {(1-m)^2 \over m^4} 
\int d^d r \: \rho \: g^{+-}(r) \sum_{\mu \nu} V^{+-}_{\mu\mu\nu\nu}(r)
+ \nn \\ &-& {\beta^2 \over 2} {(m-1) \over m^3} 
\int d^d r \: \rho \: g^{+-}(r) \sum_{\mu \nu} V^{+-}_{\mu\nu}(r) \:
V^{+-}_{\mu\nu}(r).
\eq
and the expression of $a_2^{--}$ is obtained by changing the $'+'$ in 
$'-'$ in the coefficient $a_2^{++}$.

To obtain the Legendre transform one must
solve the system of linear equations

\bq
\frac{\partial \phi}{\partial (1/ \app)} & = & -
\frac{d \: (1-m)}{2} A^+\: c_+, \nonumber \\
\frac{\partial \phi}{\partial (1/ \amm)} & = & -
\frac{d \: (1-m)}{2} A^-\: c_-,
\label{ltbb}
\eq

\n
and substitute the solutions into:

\be
\beta G(A^+,A^-,m,\beta)=\phi(\app,\amm,m,\beta)+ 
\frac{d \:(1-m)}{2} \: c_+ \frac{A^+}{\app} +
\frac{d \:(1-m)}{2} \: c_- \frac{A^-}{\amm}.
\label{lt2b}
\ee

\n
getting:

\bq
\beta G(A^+,A^-,m,\beta) & = & \gamma_0 + \gamma_3 \left (
c_+ \log A^+ + c_- \log A^- \right ) +
c_+ \: \gamma_1^+ \: A^+ + c_- \:\gamma_1^- \: A^- + \nn \\
& + & c_+ \: \gamma_2^{++} \: (A^+)^2 + c_- \:\gamma_2^{--} \: (A^-)^2
+ c_+ c_- \gamma_2^{+-} A^+ \cdot A^- 
\eq

\n
with

\be
\begin{array}{cclccl}
\gamma_0&=&c_0-a_0(1+ \log m)/m \hspace{.3in} & 
\gamma_3 & = & - a_0/m \vspace{.2in}\\
\vspace{.2in}
\gamma_1^+ & = & m \: a_1^+ \hspace{.3in} &  
\gamma_1^- & = & m \: a_1^-  \\
\vspace{.2in}
\gamma_2^{++} &=& m^2 a_2^{++}  + m^3 (a_1^{++})^2/(2 \: a_0) \hspace{.3in} &
\gamma_2^{--} &=& m^2 a_2^{--}  + m^3 (a_1^{--})^2/(2 \: a_0) \hspace{.3in} \\
\gamma_2^{+-} &=& m^2 a_2^{+-}.  &
\end{array}
\ee



\end{document}